\documentclass[notitlepage]{article}

\newcommand{\equ}[1]{~Eq.~(\ref{#1})}
\newcommand{\no}[1]{~(\ref{#1})}

\newcommand{\vev}[1]{\left\langle #1 \right\rangle}
\newcommand{\bl}[1]{\begin{equation}\label{#1}}
\newcommand{\ee}{\end{equation}}
\newcommand{\bal}[1]{\begin{eqnarray}\label{#1}}
\newcommand{\ea}{\end{eqnarray}}
\newcommand{\bx}{{\bf X}} \newcommand{\by}{{\bf Y}}

\newcommand{\bv}{{\dot{\bx}}}
\newcommand{\ba}{{\stackrel{\cdot\cdot}\atop{\hskip-0.85em\bx}}}
 \newcommand{\be}{{\bf e}}
\newcommand{\bo}{{\bf 0}}
\newcommand{\lless}{{\,<\!\!<\,}}
\renewcommand{\ll}{{|\!|}}

\newcommand{\Ai}{{\rm Ai}}
\newcommand{\G}{{\cal G}}
\newcommand{\cut}[1]{}

\begin{document}
\input{epsfig1.sty}
%\preprint{NYU-TH/99/03/xx}
%\draft

\title{Diffraction in the Semiclassical Approximation to Feynman's
Path Integral Representation of the Green Function}
\author{Martin Schaden and Larry Spruch\\New York University, Physics Department\protect\\
 4 Washington Place, New York, New York 10003}
\date{\today}
\maketitle

\begin{abstract}
We derive the semiclassical approximation to Feynman's path
integral representation of the energy Green function of a massless
particle in the shadow region of an ideal obstacle in a medium.
The wavelength of the particle is assumed to be comparable to or
smaller than any relevant length of the problem. Classical paths
with extremal length partially creep along the obstacle and their
fluctuations are subject to non-holonomic constraints. If the
medium is a vacuum, the asymptotic contribution from a single
classical path of overall length $L$ to the energy Green function
at energy $E$ is that of a non-relativistic particle of mass
$E/c^2$ moving in the two-dimensional space orthogonal to the
classical path for a time $\tau=L/c$. Dirichlet boundary
conditions at the surface of the obstacle constrain the motion of
the particle to the exterior half-space and result in an effective
time-dependent but spatially constant force that is inversely
proportional to the radius of curvature of the classical path. We
relate the diffractive, classically forbidden motion in the
"creeping" case to the classically allowed motion in the
"whispering gallery" case by analytic continuation in the
curvature of the classical path. The non-holonomic constraint
implies that the surface of the obstacle becomes a
zero-dimensional caustic of the particle's motion. We solve this
problem for extremal rays with piecewise constant curvature and
provide uniform asymptotic expressions that are approximately
valid in the penumbra as well as in the deep shadow of a sphere.
\end{abstract}

%\Large
\section{Introduction}

It was some three hundred years after Descarte's work on
diffraction theory that Keller\cite{Ke62} provided a sound
mathematical foundation for the theory by examining the asymptotic
expansion of the energy Green function. Armed with that solid
footing, and using considerable physical insight, Keller and
co-workers\cite{KL70,KeCo} and others\cite{others} found a host of
applications. Though the theory has had any number of successes,
its application can sometimes be cumbersome and its validity is
limited to the deep-shadow region. The limitation  is not really a
restriction in the asymptotic evaluation for high energies, since
any point in this case is either in the lit or the deep shadow
region of an obstacle. It does however point to the limitations of
the approach itself that have only been overcome\cite{Pr96,const}
by considering a different expansion in the penumbra.

All these approaches are purely classical, based on the eikonal-
(or ray-) approximation to the wave equation. The asymptotic forms
of the wave function and of the energy Green function thereby
obtained give the diffraction pattern for wave number
$k\sim\infty$, providing the leading correction to geometrical
optics due to diffraction. Keller interpreted his
result\cite{Ke62} geometrically in terms of paths, including those
of certain "creeping" rays, a very physical and useful way of
viewing the problem. We here study diffraction from the point of
view of a semiclassical expansion. Our starting point is the
Feynman path integral representation for the Green function of a
massless scalar field in the presence of ideal obstacles. To
simplify the presentation we assume Dirichlet boundary conditions
for the field at the surface of the obstacles and often assume a
uniform surrounding medium. We analyze and approximate the path
integral representation of the Green function semiclassically.
Note that Feynman's approach was published in 1949\cite{Fe48}, but
its advantages for obtaining asymptotic solutions -- not only
technically, but in the clearer physical picture it provides --
may not have been apparent at the time the modern version of
diffraction theory was developed. Furthermore, it seemed natural
to attack a classical theory classically. In fact,
Gutzwiller\cite{Gu90} and others have made significant advances in
the semiclassical evaluation of the path integral. To our
knowledge these advantages of Feynman's formulation and
Gutzwiller's semiclassical evaluation have never been fully
exploited in the case of diffraction theory.  (Some related
approaches by others will be briefly commented on below.) We here
hope to help close this gap.

That an asymptotic analysis of the classical wave equation for
large $k$ should give the same result as the semiclassical
expansion is well known but may nevertheless warrant a remark. For
simplicity, we assume for the moment that the medium is uniform.
Since the connection between the local wave number $k$ and the
energy $E$ of a massless particle is
\bl{defk}
k=E/(\hbar v(E))\ ,
\ee
where $v(E)$ is the phase velocity in the medium, letting $k$ be
arbitrarily large is equivalent to letting $\hbar$ be arbitrarily
small.

Our final result is equivalent to those obtained previously in
certain limits. However, the present development may have a number
of conceptual advantages. The illuminated region, the penumbra and
the umbra are all treated in a unified fashion. The approach
furthermore is probably more accessible to most physicists than
the original one, and might provide new insights. By providing a
different point of view of the phenomenon, it may also suggest
other (interpolating) approximations. Perhaps the most interesting
aspect of the present formulation is that it provides a systematic
geometrical framework that in principle allows one to obtain
asymptotic expressions for diffraction for many systems, although
considerable numerical work may be necessary if the geometrical
setting is sufficiently complicated; in particular, it shows that
diffraction can be understood as a special case of quantum
mechanical tunnelling in the sense that the corresponding motion
is classically forbidden and that the semiclassical approximation
is obtained by analytic continuation in some of the parameters of
the problem.

We will study the semiclassical approximation,
$G(\by,\bx;E)$, to the energy Green function
\bl{GreenE}
\G(\bx,\by;E)\equiv \vev{\by|\hat{\cal G}(E+i\eta)
|\bx}=\vev{\by|(\hat H-E-i\eta)^{-1}|\bx}\ ,
\ee
the coordinate representation of the Laplace transform of the time
evolution operator. For simplicity we will study only the
propagation of  a free massless scalar particle in a medium which,
for the most part, will be taken to be homogeneous. We require
that $\G$ vanish for $|\bx-\by |\rightarrow \infty$ and also
whenever $\bx $ or $\by $ are on a smooth compact (but not
necessarily connected) two-dimensional surface ${\cal F}$. These
Dirichlet boundary conditions are also the appropriate ones for a
particular polarization of the electromagnetic field in the
presence of ideal conductors. Note that semiclassically only
photons whose polarization is normal to the surface can diffract
in the case of an ideal conductor.

Green functions are among the most basic concepts in physics.
Thus, for example, the trace of $\hat {\cal G}(E+i\eta)$ is the
response function $g(E)$, whose imaginary part gives the spectral
density $\varrho(E)$, where $\varrho(E) dE$ is the number of
states with energy between $E$ and $E+dE$. Unfortunately, Green
functions can rarely be obtained exactly. An exceedingly useful
result, largely due to Gutzwiller\cite{Gu90}, is the observation
that the semi-classical approximation to ${\cal G}(\by,\bx ;E)$
and $g(E)$ are usually completely determined by the solution of a
corresponding {\it classical} problem. Unfortunately, the solution
must be modified if it is to be used in the case of diffraction.
The corresponding classical trajectories are {\it non-holonomic}
constrained extrema of the action that are not stationary; the
dependence of the action on the fluctuations therefore contains
linear terms. A naive application of the usual semiclassical
approximation fails. We will see that the semiclassical Green
function $G(\by,\bx;E)$ nevertheless {\it is} given by the
solution to a corresponding classical problem in this case too.
The corresponding classical problem one has to solve in the case
of diffraction is that of the motion of a non-relativistic
particle in two dimensions under the influence of a generally
time-dependent gravitational-like acceleration inversely
proportional to $R(s(t))$ in the presence of a ceiling; $R(s(t))$
is the radius of curvature of the constrained {\it classical}
trajectory connecting $\bx$ and $\by$ as a function of the arc
length $s(t)$.

Our interest in semiclassical diffraction theory was rekindled in
the course of a study of Casimir effects. We found that a
semiclassical evaluation gives the leading behavior whenever the
Casimir energy diverges \cite{us} as one length scale of the
problem is taken to be much larger than any other. In many cases
this asymptotic evaluation is sufficient to determine the Casimir
energy {\it exactly}. A phenomenologically interesting example is
Derijaguin's problem\cite{De68} of the Casimir force between two
conducting spheres of radii $R_1$ and $R_2$ in the limit where
their separation $d$ is arbitrarily small compared to $R_1$ and
$R_2$. This force had been determined using very plausible
arguments\cite{De68,Mo97}, but the semi-classical calculation
provided what we believe to be the first rigorous derivation of
this result.  The Casimir force in this limiting case is
proportional to  $\bar R/d^3$, where $\bar R=R_1 R_2/(R_1+R_2)$,
and diverges as $\bar R/d\rightarrow \infty$; the semiclassical
calculation gives the correct coefficient. On the other hand, for
separations large compared with the radius of either sphere, the
Casimir-Polder force\cite{CasimirPolder} falls off as $1/d^8$ and
{\it a priori} one cannot assert that the semiclassical
approximation must be exact in this case. To see whether this
approximation qualitatively reproduces the $1/d^8$ behavior of the
force in this limit, one has to take diffraction into account at
least semiclassically. The effort to include these effects and
thus perhaps extend the validity of the semiclassical calculation
beyond the regime $d\lless\bar R$ led us to the present study. The
extension to
$d{\textrm{\raisebox{-.3ex}{$>$}}\atop\textrm{\raisebox{.1ex}{$\sim$}}}
\bar R$ is under consideration.

In Sec.~II we present a heuristic argument based on Fermat's
principle to derive the effective action for the semiclassical
approximation to the path-integral representation of the energy
Green function of a massless particle in the presence of idealized
obstacles. There is no stationary classical path that can describe
propagation into the shadow region of the obstacle. The situation
turns out to be analogous to that of calculating the amplitude
that a non-relativistic particle move in finite time from the
ceiling to the ceiling under the influence of a gravitational
force. We show in Sec.~III that an analytic continuation in the
parameters solves the problem by changing it to that of a
non-relativistic particle moving in finite time from the floor to
the floor under the influence of a gravitational-like force -- the
same problem encountered in the semiclassical description of
"whispering galleries". The remainder of Sec.~III is devoted to an
analytical study of diffraction by a sphere. In Sec.~IV we relate
the description by classical trajectories in a gravitational-like
field to Keller's asymptotic evaluation of the energy Green
function for this example. In particular, we also derive the power
corrections of the asymptotic expansion within this unified
approach.

There is an immense literature on the subject. See, for example
Gutzwiller\cite{Gu90} and Reichl\cite{Re92}, and references
therein. We also include a small sampling of recent developments
on caustics\cite{lit}.

\section{The Action: A Heuristic Argument}
Our ultimate goal is to obtain the  semiclassical approximation
${G}(\bx,\by;E)$  to $\G(\bx,\by;E)$, the probability amplitude
that a photon of energy $E$ at the initial point $\bx$ will at
some time appear at $\by$. Since the semiclassical energy Green
function is essentially determined by the quadratic action
$S^{sc}$ for the fluctuations -- as we will discuss later -- we
first determine the form of this action. (See\equ{semi2} below.)

\subsection{The Action for the Fluctuations to Quadratic Order}
We argue that the contribution to the energy Green function of a
continuous path $\gamma$ from $\bx$ to $\by$ is determined by the
action
\bl{Fermat}
S(\gamma,E)=E \tau(\gamma,E)=E \int_0^T dt
|\dot\bx(t)|/v(\bx(t),E)\ .
\ee
A path $\gamma$ from $\bx$ to $\by$ in this context is defined by
the three continuous functions on the interval $0\leq t\leq T$
that make up the vector $\bx(t)$ and satisfy $\bx(0)=\bx$ and
$\bx(T)=\by$. The variable $t\in [0,T]$ need not be the actual
time, since the integral in\equ{Fermat} does not depend on how the
path is parameterized. Using the arc length $s$ along the path
$\gamma$, one may rewrite $S(\gamma,E)$ as
\bl{arcS}
S(\gamma,E)=E\int_0^{L_\gamma} \frac{d s}{v(s,E)}\ ,
\ee
where $L_\gamma$ is the path's total length. For the semiclassical
quantization it will be important that \equ{Fermat} in fact does
not depend on the "speed" $|\dot \bx|$, subject to the restriction
that $|\dot \bx|$ never vanish along the path (or equivalently
that the arc length is a monotonically increasing function of the
parameter $t$). $2\pi\nu\tau(\gamma,E)$ is the phase lag of a
monochromatic wave of frequency $\nu=E/(2\pi\hbar)$ along the path
$\gamma$ with the phase velocity $v(s,E)$ related to the {\it
local} index of refraction $n(s,\nu)=c/v(s,E)$ of the medium along
the path. Fermat's principle is the statement that the
``classical'' trajectory $\gamma_c$  of a monochromatic ray is an
extremum of $\tau$. It is the phase velocity that determines the
{\it trajectory} of a monochromatic ray (as is for instance
evident from Snell's law). Other than when the medium is a vacuum,
the time $\tau(\gamma_c, E)$ is {\it not} the time between
emission at $\bx$ and absorption at $\by$ of a {\it photon} of
frequency $\nu$. The travel time for a photon or, classically, a
wave-packet, is determined by the local group velocity $v_g(s,E)$
along the path,
\bl{groupv}
v^{-1}_g(s,E)=\frac{\partial}{\partial
E}\left(\frac{E}{v(s,E)}\right)\ ,
\ee
that is, the time ${\cal T}$ for a photon of energy $E$ to travel
from $\bx$ to $\by$ is
\bl{tott}
{\cal
T}=\int_0^{L_\gamma}\frac{ds}{v_g(s,E)}=\frac{\partial}{\partial
E}\int_0^{L_\gamma} ds \frac{E}{v(s,E)} =\frac{\partial
S(\gamma,E)}{\partial E}\ .
\ee
The relation\equ{tott} between the total time ${\cal T}$ and the
quantity $S$ defined by\equ{Fermat}, in conjunction with Fermat's
principle, identifies $S(\gamma,E)$ as the classical action that
describes the motion of a massless particle (uniquely up to an
irrelevant constant that does not depend on the energy nor on the
path $\gamma$). Hamilton's principal function $R(\bx,\by; {\cal
T})$ with the independent variable ${\cal T}$ is the Legendre
transform of $S$,
\bl{HPF} R(\bx,\by;{\cal T})=S(\bx,\by,E)-E{\cal T}=\left.E^2
\frac{\partial}{\partial E}\int_0^{L_\gamma}
\frac{ds}{v(s,E)}\right|_{E\rightarrow E({\cal T})}
\ee
and vanishes in a vacuum, as it must, because the Minkowski
distance between the events vanishes and Hamilton's principal
function is proportional to it.

We temporarily restrict our considerations to the situation where
the phase velocity $v(\bx(t),E)=v(E)$ is  constant when $\bx(t)$
is exterior to the volume ${\cal V}$ with surface ${\cal F}$ and
vanishes when $\bx(t)$ is in the interior of ${\cal V}$. This is
an idealized limit of the physical situation where obstacles with
a very high index of refraction are embedded in a homogeneous
medium. The index of refraction in the physical case depends
smoothly on the coordinates and the classical paths are stationary
points of the action. The action defined by\equ{Fermat} of a path
that avoids ${\cal V}$ in going from $\bx$ to $\by$ is
proportional to its length. The vanishing phase velocity in the
interior of ${\cal V}$ leads to a non-holonomic constraint on the
classical motion. We will see that the situation is akin to motion
in a gravitational-like potential in the presence of a ceiling.
The classical motion is still an extremum of the action, simply
because the constraint can be viewed as a particular {\it limit}
of the physical case. In the limit, the extremum generally is not
a stationary point of the action. The change of the action under
some small deviations from the extreme path can no longer be made
arbitrarily small in this limit, that is, the functional
derivative of the action at the extremum does not exist.
{\vskip0.1truecm\epsfig{figure=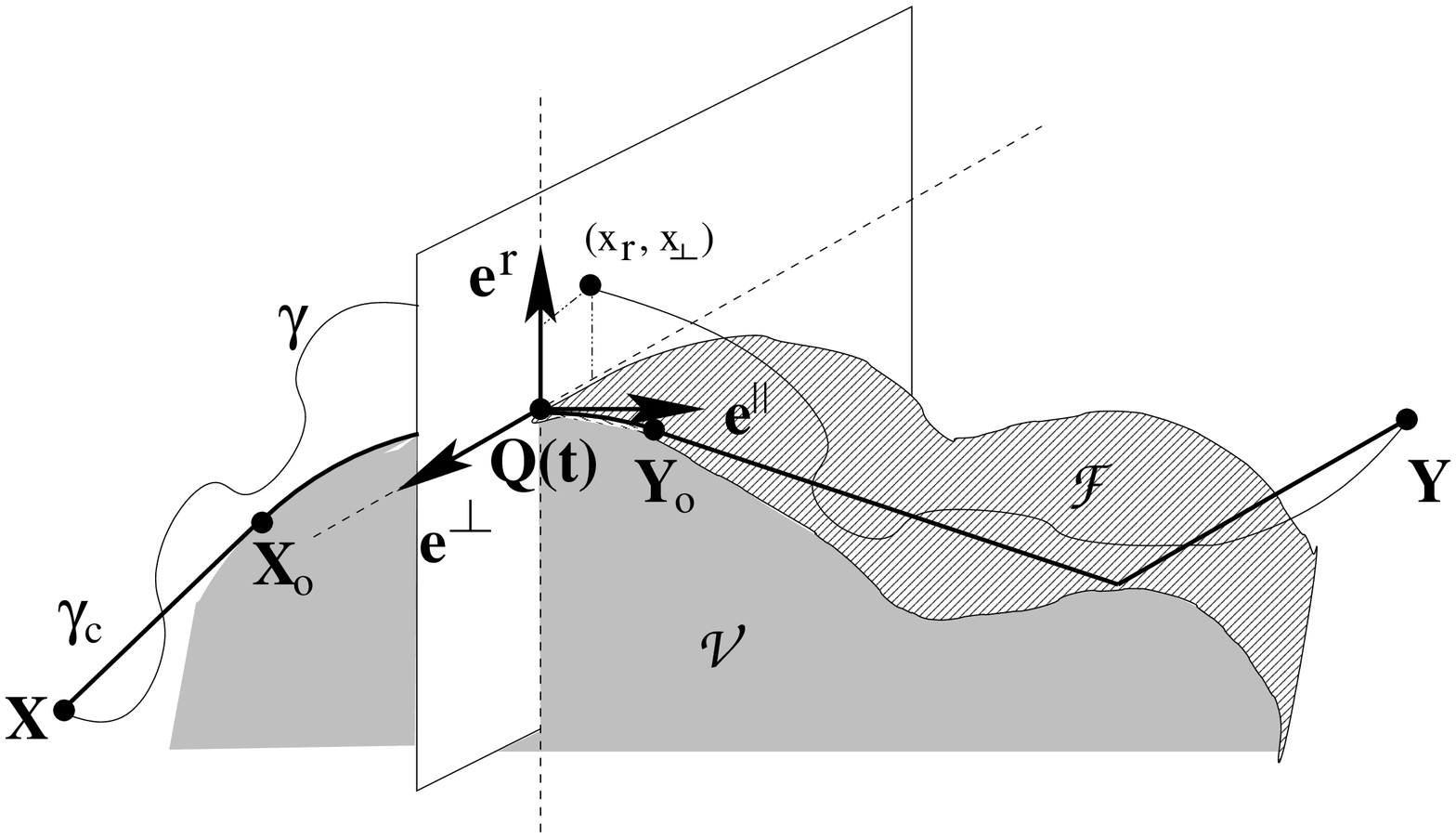,height=6.0truecm}\nobreak

{\small\noindent Fig.~1: The semiclassical expansion. The bold
curve is a schematic plot of a classical path $\gamma_c$ of
extremal length from $\bx$ to $\by$ that is excluded from the
(smooth) volume ${\cal V}$. The length of $\gamma_c$ is extremal
but not stationary, since it "creeps" along the surface between
the points $\bx_o$ and $\by_o$. A non-extremal path $\gamma$ in
the vicinity of this classical path is shown as a thin line and
the local coordinate frame at the point ${\bf Q}(t)$ of $\gamma_c$
is sketched. The fluctuation $\gamma$ is fully described by the
coordinates $x_r(t)$ and $x_\perp(t)$ of local coordinate frames
along the whole classical path. Note that $\be^\perp$ lies in the
surface ${\cal F}$.}

As can be seen from Fig.1, the classical trajectory of extremal
length from $\bx$ to $\by$ can be rather complicated even in the
presence of smooth obstacles. It may consist not only of straight
sections and reflections off points on the surface, but also of
segments that ``creep'' along the surface. The classical
trajectory is by definition continuous and of extremal length in
coordinate space. It is natural to parameterize it by the
``intrinsic'' time $t$ for which $s(t)= v(E) t$ is the arc length
along the classical path. In this case the instantaneous velocity
$\bv_c(t)$ and the instantaneous acceleration $\ba_c(t)$ are
orthogonal vectors, since $|\bv_c(t)|=v(E)$ does not depend on
time.

This singles out a natural local orthonormal coordinate frame at
each point along curved sections of the classical path. The basis
vectors of this local coordinate frame are
\bl{locbasis}
\be^\ll(t)=\frac{\bv(t)}{v(E)}\ ,\ \ \be^r(t)=- \frac{R(t)}{v^2(E)}\,\ba(t) \ ,\ \
\be^\perp(t)=\be^\ll(t)\times\be^r(t)\ ,
\ee
where the local radius of curvature of the classical path is
related to the instantaneous acceleration by
$R(t)=v^2(E)/|\,\ba(t)|$. $\be^\ll(t)$ clearly is parallel to the
velocity $\bv(t)$ and thus tangent to the classical path,
$\be^r(t)$ lies in the local plane of motion (and is normal to the
surface of the obstacle) while $\be^\perp(t)$ is perpendicular to
the local plane (and lies in the surface of the obstacle). On
straight sections of the classical path a similar local
orthonormal coordinate frame exists, but is not uniquely defined
with respect to rotations about the classical path.

We next introduce local coordinates that give the deviation of the
path $\gamma$ from a classical path $\gamma_c$ that is an extremum
of the action in\equ{Fermat}. Let $\bx_c(t)$ and $\bx(t)$ describe
a point on the classical trajectory $\gamma_c$ and a point on the
path $\gamma$
 in a (time-independent) global coordinate frame at the intrinsic time $t$
of the classical path. The coordinates $x_\ll(t),x_r(t)$ and
$x_\perp(t)$ for the fluctuations in the local basis at time $t$
are then defined by
\bl{fluc}
\bx(t)=\bx_c(t)+ x_\ll(t)\be^\ll(t) +x_r(t)\be^r(t) +
x_\perp(t)\be^\perp(t)\ .
\ee
where $\bx(t)$ is a point on $\gamma$. (See Fig.~1). It is
important to note that certain fluctuations just reparameterize
the classical path. This follows from our earlier observation that
the speed along the path plays no role. More formally, if instead
of the intrinsic time $t$ one had chosen $t^\prime$, where
\bl{repar}
t^\prime=t+\eta(t)\ ,
\ee
to parameterize the classical path, then to first order in $\eta(t)$
\bl{repar1}
\bx_c(t^\prime)= \bx_c(t)+\eta(t)\bv_c(t) +\dots =\bx_c(t)
+\eta(t)v(E)\be^\ll(t)+\dots
\ee
An infinitesimal longitudinal fluctuation, $x_\ll(t)=\eta(t)v(E)$,
thus describes the {\it same} classical path and is equivalent to
the use of a slightly different ``clock'' to parameterize the
classical path. To avoid this ambiguity and have a one-to-one
correspondence between the fluctuations and paths, one considers
only fluctuations in the planes
\bl{gauge}
x_\ll(t)=0\ .
\ee
We can do so, because the local coordinates $x_r(t)$ and
$x_\perp(t)$ in each plane along the classical path $\gamma_c$
completely specify $\gamma$. The geometrical meaning and
uniqueness of this construction can be seen in Fig.~1.

We will also need to know how the basis vectors defined by
\equ{locbasis} rotate with time. Since, by\equ{locbasis},
\bl{rotparallel}
\dot\be^\ll(t)=\ba(t)/v(E)=-v(E)\be^r(t)/R(t)\
\ee
is in the direction of one of the basis vectors, the rotation of
the coordinate frame with time is specified by only two (generally
time-dependent) angular velocities, instead of the usual three.
The basis at time $t$ is related to the basis at time $t+dt$ by an
infinitesimal orthogonal transformation $\Omega$. $\be^\ll(t+dt)$
thus is given in terms of the elements of $\Omega$ and of the
$\be(t)$'s of\equ{locbasis}. Comparing the resulting  expression
with\equ{rotparallel} determines two of the three independent
elements of $\Omega$. Denoting the as yet to be determined
independent element of $\Omega$ by $\alpha(t)$, one obtains
\bl{dotbasis}
\dot\be^\ll(t)=-\frac{v(E)}{R(t)}\be^r(t)\ ,\ \
\dot\be^r(t)=\frac{v(E)}{R(t)}\be^\ll(t) - \alpha(t)\be^\perp(t)\
,\ \ \dot\be^\perp(t)=\alpha(t)\be^r(t)\ .
\ee
Explicitly differentiating $\be^r(t)$ in\equ{locbasis} and taking
the inner product with $\be^\perp(t)$ determines $\alpha(t)$ as
\bl{skew}
\alpha(t)=\frac{R(t)}{v^2(E)}
({\stackrel{...}\atop{\!\!\!\!\!\!\bx_c}}(t)\cdot\be^\perp(t))\ .
\ee
$\alpha(t)$ vanishes if the classical trajectory lies in a plane
and thus is a measure of its skewness. It is the rate at which the
coordinate frame rotates about the tangent vector to the classical
path. For a helix on a cylinder of radius $R_{\rm cyl}$, a
velocity $v_z$ parallel to the cylinder axis and an angular
velocity $\omega$, one finds $\alpha(t)=-\omega
v_z/\sqrt{v_z^2+(\omega R_{\rm cyl})^2}$, independent of
time\cite{fnote1}.

We are now in a position to expand the action of\equ{Fermat} to quadratic
order in the fluctuations $x_\ll=0, x_r(t),x_\perp(t)$ that uniquely
describe a path in the vicinity of the classical one.  Taking the time
derivative of\equ{fluc} and using\equ{dotbasis} to describe the time
dependence of the basis one finds
\bal{vfluc}
\bv(t)&=&\left[\left(\dot x_r(t)+\alpha(t) x_\perp(t)\right)
\be^r(t) +\left(\dot x_\perp(t)-\alpha(t)
x_r(t)\right)\be^\perp(t)\right]\cr
&&\quad+v(E)\left(1+\frac{x_r(t)}{R(t)}\right)\be^\ll(t)\ .
\ea
As noted earlier $v(E)\be^\ll(t)$ is the velocity $\bv_c(t)$ along
the classical trajectory. [\equ{locbasis} gives the same
relationship, but with $\bx_c(t)$ rather than $\bx(t)$. There is
no inconsistency, since we have chosen $x_\ll(t)=0$.] Expanding
the action of\equ{Fermat} to quadratic order in the fluctuations
using\equ{vfluc}, one finally obtains
\bal{semi2} S^{sc}_{\gamma_c}(\gamma,E)&=&\int_0^{\tau(\gamma_c,E)}
 \!\!\!\!\!\!\! dt \left\{E+E\frac {x_r(t)}{R(t)}\right.\cr
 &&\hskip-2em+\left.\frac{E}{2v^2(E)}\left[\left(\dot x_r(t)+
 \alpha(t)x_\perp(t) \right)^2
+\left(\dot x_\perp(t)-\alpha(t)x_r(t)\right)^2\right]\right\}\
\cr &&
\ea
for the action of the path $\gamma$ in semiclassical
approximation, when the classical path is $\gamma_c$. Note that
the term {\it linear} in the fluctuations of the semiclassical
action\no{semi2} is due to the fact that the classical path,
constrained to lie outside of ${\cal V}$, is extremal, but not
stationary. Since the classical trajectory described by
$x_r(t)=x_\perp(t)=0$ creeps along the surface ${\cal F}$ whenever
$0<R(t)<\infty$ and since $\be^r(t)$ is normal to ${\cal F}$ at
that point,  the coordinate $x_r(t)$ takes only positive values on
a creeping segment. The kinetic terms in\equ{semi2} are those of a
non-relativistic particle of mass
\bl{mass}
m_E=E/v^2(E)
\ee
moving in a plane that is rotating with angular velocity
$\alpha(t)$. To explicitly see this, we perform a time-dependent
orthogonal transformation of the coordinates
\begin{eqnarray}\label{transform}
x_r(t)&=&{\underline x}_r(t) \cos\theta(t) - {\underline
x}_\perp(t) \sin\theta(t)\cr x_\perp(t)&=&{\underline x}_r(t)
\sin\theta(t)+{\underline x}_\perp(t) \cos\theta(t)
\end{eqnarray}
with
\bl{angle}
\theta(t)=\int_0^t dt^\prime \alpha(t^\prime)\ .
\ee
The expressions in\equ{semi2} are then
\bal{velocities}
\dot x_r(t)+\alpha(t)x_\perp(t)&=&\dot{{\underline x}}_r(t)\cos
\theta(t)-\dot{{\underline x}}_\perp(t) \sin\theta(t)\cr \dot
x_\perp(t)-\alpha(t)x_r(t)&=&\dot{{\underline x}}_r(t)\sin
\theta(t)+\dot{{\underline x}}_\perp(t) \cos\theta(t)\ .
\ea
Written in terms of ${\underline x}_r(t)$ and ${\underline
x}_\perp(t)$ the semiclassical action of\equ{semi2} becomes
\bl{semi3}
S_{\gamma_c}^{sc}(\gamma)=\int_0^{\tau(\gamma_c,E)}\!\!\!\!\!\!
dt\, {\cal L}_{\gamma_c}({\underline x}_r, \dot{{\underline
x}}_r,{\underline x}_\perp,\dot{{\underline x}}_\perp;E,t)\ ,
\ee
where the two-dimensional Lagrangian,
\bal{L}
{\cal L}_{\gamma_c}({\underline x}_r, \dot{{\underline
x}}_r,{\underline x}_\perp,\dot{{\underline x}}_\perp;E,t)&=& E
+\frac{m_E}{2}\left(\dot{{\underline x}}_r^2(t) + \dot{{\underline
x}}_\perp^2(t)\right)\cr && - V_E\left[{\underline x}_r(t)
\cos\theta(t) - {\underline x}_\perp(t) \sin\theta(t);t\right]\
,\cr &&
\ea
contains an explicitly time-dependent potential term
\bl{pot}
V_E[z;t]=E\left\{
\begin{array}{cll}
-z/R(t) &,&{\rm if}\ z\ge 0\ {\rm and}\ 0<R(t)<\infty\cr \infty
&,&{\rm if}\ z<0\ {\rm and}\ 0<R(t)<\infty\cr 0 &,&{\rm
otherwise}\ ,
\end{array}\right.
\ee
that vanishes at times $t$ for which the classical path does not
lie on the surface ${\cal F}$.

\subsection{The Semiclassical Green function}
The semiclassical action of\equ{semi3} can be interpreted as
describing the non-relativistic motion of a particle of mass
$m_E=E/v^2(E)$ in a homogeneous but time-dependent
gravitational-like potential in two dimensions. To complicate
matters there is a (generally time-dependent) restriction that
$x_r(t)={\underline x}_r(t) \cos\theta(t) + {\underline
x}_\perp(t) \sin\theta(t)\ge 0$ whenever $0<R(t)<\infty$. In
analogy to motion in a gravitational field, the restriction
$x_r(t)\ge 0$ will be referred to as due to a (time-dependent)
{\it ceiling}.  Since the force is always directed {\it away} from
the ceiling, there is {\it no classical trajectory} that starts
from the ceiling and ends at the ceiling after a finite time
$\tau(\gamma_c,E)$. However, the quantum mechanical amplitude for
again observing the particle  at the ceiling after a time
$\tau(\gamma_c,E)$ does not vanish. We now proceed to calculate
this amplitude and thus describe diffraction semiclassically.

To simplify the notation, let us introduce the two-dimensional
position- and momentum- vectors
\bl{vecxp}
{\underline{\vec x}}=({\underline x}_r,{\underline x}_\perp),\ \
{\underline{\vec \pi}}=(m_E\dot{{\underline
x}}_r,m_E\dot{{\underline x}}_\perp)\
\ee
to describe the motion of the  particle in phase space. The
semiclassical Hamiltonian $H_{\gamma_c}$ of our two-dimensional
problem depends on the classical path $\gamma_c$ and is explicitly
time dependent. It is given by a Legendre transformation of the
Lagrangian of\equ{L} and in the notation of\equ{vecxp} reads
\bl{ham}
H_{\gamma_c} ({\underline{\vec x}},{\underline{\vec \pi}}; t)
={\underline{\vec \pi}}\cdot{\underline{\dot{\vec x}}} -{\cal
L}_{\gamma_c}= \frac{{\underline{\vec \pi}}^{\,2}}{2 m_E} +
V_E(x_r({\underline{\vec x}},t);t) -E .
\ee
Note that the {\it parameter} $E$ of this two-dimensional
Hamiltonian also determines the zero of the energy scale. This
constant is unimportant for the dynamics at a {\it given} value of
$E$,  but is relevant when comparing the phases of the amplitudes
at different wave numbers. Upon quantization of the classical
problem described by the Hamiltonian of\equ{ham}, the amplitude
that a particle in this two-dimensional world found initially at
the transverse deviation  ${\underline{\vec x}}=\vec a$ from the
classical path will appear with transverse deviation
${\underline{\vec x}}=\vec b$ a time $\tau(\gamma_c,E)$ later is
the causal Green function
\bal{Greentau}
\G_{\gamma_c}(\vec a,\vec b;\tau(\gamma_c,E)\, )&=& \langle\vec b|
\exp\left\{-\frac{i}{\hbar} \int_0^{\tau(\gamma_c,E)}dt\hat
H_{\gamma_c}({\underline{\vec x}},{\underline{\vec \pi}};
t)\right\} |\vec a\rangle\cr &=&\int_{\gamma } [dx] \exp
\frac{i}{\hbar} S_{\gamma_c}^{sc}(\gamma) \ .
\ea
The (time-ordered) exponential in\equ{Greentau} is the
time-evolution operator generated by a time-dependent Hamiltonian
for a fixed time interval -- here $\tau(\gamma_c,E)$, for given
$E$ -- and the paths $\gamma$ in the latter representation of the
Green function as a path integral in coordinate space are
described by their transverse deviations,  $\vec a$ at $t=0$  and
$\vec b$ at $t=\tau(\gamma_c,E)$, from the classical path
$\gamma_c$. (We use the notation $\G$ and $G$ for exact and
semiclassical Green functions. Rather than introduce a new symbol,
we choose to write $\G_{\gamma_c}$ for the exact Green function to
the semiclassical Hamiltonian of\equ{ham} for the classical path
$\gamma_c$.)

The semiclassical energy Green function $G(\bx,\by;E)$ is
proportional to the sum over all classical trajectories $\gamma_c$
of the amplitudes, $\sum_{\gamma_c} \G_{\gamma_c}(\vec a=\vec
b=0;\tau(\gamma_c,E)\, )$. As noted in Appendix~A, the two Green
functions are related\cite{Gu90a} by a factor that depends only on
the local {\it group}  velocities $v_g(\bx,E)$ and $v_g(\by,E)$ at
the endpoints of the classical trajectories,
\bl{rel1}
G(\bx,\by;E)=\frac{1}{i\hbar\sqrt{v_g(\bx,E)
v_g(\by,E)}}\sum_{\gamma_c} \G_{\gamma_c}(\vec a=\vec
b=0;\tau(\gamma_c,E)\, )\ .
\ee
For a homogeneous medium, the group velocity in fact does not
depend on the location outside ${\cal V}$ and
$v_g(\bx,E)=v_g(\by,E)=v_g(E)$ and \equ{rel1} simplifies
accordingly.

\section{An Example: Diffraction by a Sphere in a Uniform Medium}
To better understand the general case of\equ{Greentau}, it is
illustrative to consider the example of diffraction when the
excluded volume ${\cal V}$ is a sphere of radius $R$. The exact
expressions for the Green functions are known for this case and
their asymptotic expansion for large $k R =E R/(\hbar v(E))$ has
been obtained\cite{Ke62,Pr96,const}. Besides making contact with
previous work, the spherical case also exhibits some analytical
properties of the Green function $\G_{\gamma_c}$ that are
essential in the construction of the solution for a more general
obstacle.
{\vskip0.1truecm\epsfig{figure=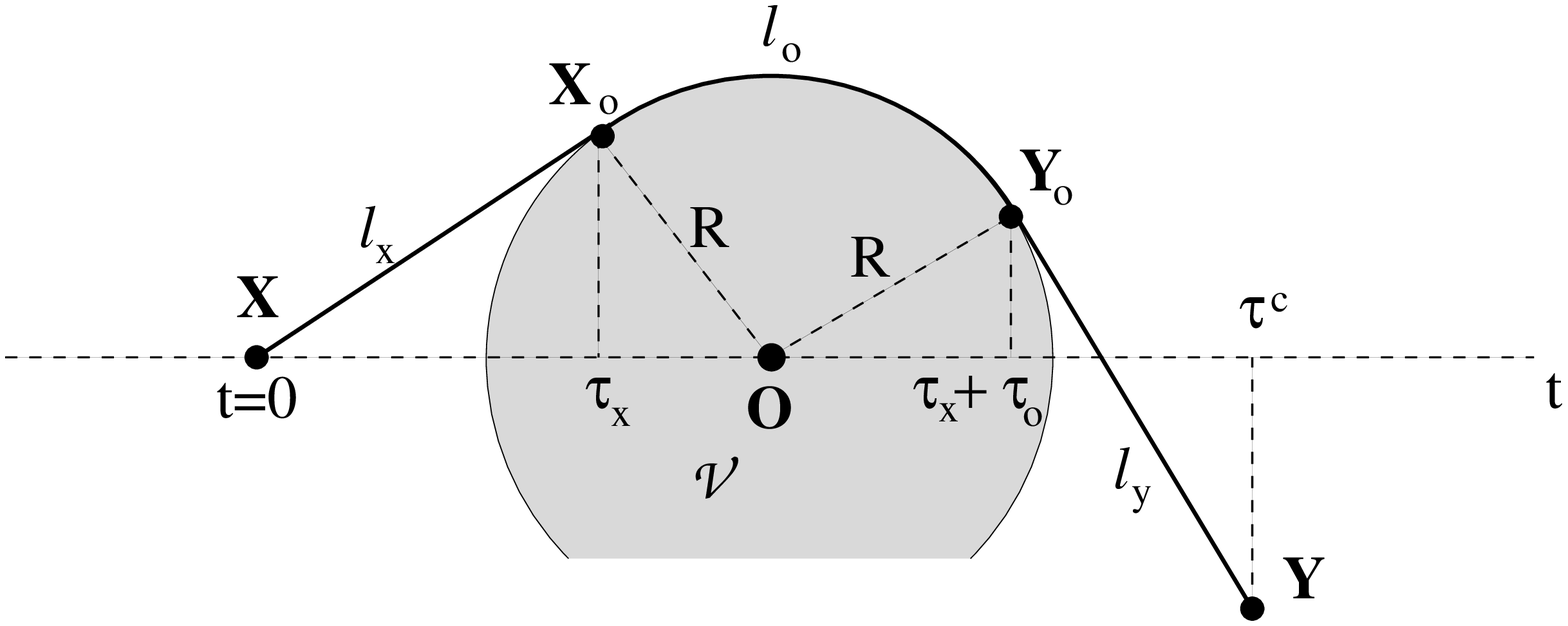,width=11.0truecm}\nobreak

{\small\noindent Fig.~2:The classical path, from ${\bx}$ at $t=0$
to $\by$ at $t=\tau_c$ with winding number $w=+1$, for a spherical
excluded volume ${\cal V}$ of radius $R$. The path is planar and
$l_x$ and $R$ fix the location of $\bx$ with respect to the center
of the sphere, $\bo$, and $l_o$ and $l_y$ then determine $\by$.
Note that for $|w|>1$, the classical path would completely
encircle the sphere. The phase velocity is $v(E)$, and
$\tau_x=l_x/v(E), \tau_o=l_o/v(E), \tau_y=l_y/v(E)$, with
$\tau_c=\tau_x+\tau_o+\tau_y$.}

The initial point $\bx$, the final point $\by$ and the center
$\bo$ of the sphere ${\cal V}$ define a plane. Every classical
trajectory lies wholly in this plane and the angle $\alpha(t)$
vanishes. As is evident from Fig.~2, classical trajectories in
this case can be assigned an integer $w=0,\pm 1,\pm 2\dots$ that
represents the number of times and the direction that the
classical trajectory winds around the sphere in going from $\bx$
to $\by$. $w=0$ corresponds to the direct trajectory and does not
occur if, for given $\bx$, $\by$ lies in the shadow cast by the
sphere. $w=\pm 1$ are the two trajectories of opposite winding
sense that do touch the sphere but do not wind fully around it,
$w=\pm 2$ is assigned to the two trajectories that wind once but
not twice around the sphere, etc\dots. In addition to the sense in
which it winds around the sphere, a classical trajectory
$\gamma_c$ is characterized by three lengths, $l_x,l_o$ and $l_y$.
$l_x$ is the length of the straight line segment of the trajectory
that is tangent to the sphere and extends from $\bx$ to a point
$\bx_o$ on the surface of the sphere. Similarly $l_y$ is the
length of the straight line segment of the trajectory  that is
tangent to the sphere and extends from $\by$ to a point $\by_o$ on
the surface of the sphere. Finally $l_o$ denotes the length of the
classical trajectory that ``creeps'' along the surface of the
sphere between $\bx_o$ and $\by_o$. The time intervals
corresponding to these lengths are $\tau_x=l_x/v(E)$,
$\tau_y=l_y/v(E)$ and $\tau_o=l_o/v(E)$, respectively, with
\bl{tcdef} \tau_c=\tau(\gamma_c,E)=\tau_x+\tau_o+\tau_y\ .
\ee
It is not difficult to ascertain that such a trajectory is of
extremal length.

The Green function $\G_{\gamma_c}({\vec a}={\vec b}=0;\tau_c)$ is
a function of the parameters $E,v(E),R$ and of the variables
$\tau_x,\tau_y,\tau_o$ describing the classical path. Since
$\alpha(t)=0$, it follows that ${\underline x}_r=x_r$ and that
${\underline x}_\perp=x_\perp$. The potential in\equ{ham}
therefore does not depend on $x_\perp$. The hamiltonian
of\equ{ham} in this case is of the form
\bl{hamsphere}
H_{\gamma_c}=H_\perp(\pi_\perp) +H_r(x_r,\pi_r;t)-E\ ,
\ee
and the dynamics of the perpendicular and radial degrees of freedom
separates.  The time-independent Hamiltonian
\bl{hamperp}
H_\perp(\pi_\perp)=\frac{\pi_\perp^{2}}{2 m_E}
\ee
describes the one-dimensional motion of a free non-relativistic
particle of mass $m_E=E/v^2(E)$, whereas
\bl{hamr}
H_r(x_r,\pi_r;t)=\frac{\pi_r^{2}}{2m_E} + V_E(x_r;t)
\ee
governs the radial dynamics. The boundary conditions $\vec a=\vec
b=0$ imply that $x_r(0)=x_r(\tau_c)=0$ and
$x_\perp(0)=x_\perp(\tau_c)=0$. By virtue of the form of
\equ{hamsphere}, the Green function $\G_{\gamma_c}({\vec a}={\vec
b}=0;\tau_c)$ decomposes into the product of two Green functions
and an exponential, and we can write
\bl{Greenr}
\G_{\gamma_c}({\vec 0},{\vec 0};\tau_c)= \exp(i E \tau_c/\hbar)
\G_0(0,0;\tau_c) \G_r(0,0;\tau_c)\ ,
\ee
where, for arbitrary $x$, $y$ and $\tau$,
\bl{G0}
\G_0(x,y;\tau)=\sqrt{\frac{m_E}{2 \pi i \hbar\tau}}
\exp\left(\frac{i m_E(x-y)^2}{2\hbar \tau}\right)
\ee
is the Green function for a free non-relativistic particle of mass
$m_E$ in one dimension and
\bl{Gr}
\G_r(x,y;\tau)=\langle y| \exp\left\{-\frac{i}{\hbar}
\int_0^{\tau} dt\hat H_r(x,\pi; t)\right\} |x\rangle
\ee
is the Green function corresponding to the time-evolution
generated by the radial Hamiltonian of\equ{hamr}. The exponential
factor in\equ{Greenr} originates in the $-E$ term
in\equ{hamsphere} and is the phase associated with the classical
path.  The explicit time dependence of the radial Hamiltonian in
the case of a sphere is quite simple since the curvature of a
classical path $\gamma_c$ is of the form
\bl{rtime}
\frac{1}{R(t)}=\left\{
\begin{array}{cll}
1/R &,& \tau_x<t<\tau_x+\tau_o\cr 0 &,& {\rm otherwise}
\end{array}\right.\ .
\ee
In the time intervals $[0,\tau_x]$ and
$[\tau_x+\tau_o,\tau_x+\tau_o+\tau_y]$ the Hamiltonian
of\equ{hamr} describes the time evolution of a free
non-relativistic particle of mass $m_E$. If the radial coordinate
states were complete, ${\cal G}_r$ could be further decomposed as
\bal{decompGr}
\G_r(0,0;\tau_c)&=&\int_0^\infty\!\!\! da\int_0^\infty\!\!\! db\,
\G_0(0,a;\tau_x) \G_R(a,b;\tau_o) {\cal G}_0(b,0;\tau_y)\cr
&&\hskip-3em=\frac{k}{2\pi i} \int_0^\infty \!\!\!
\frac{da}{\sqrt{l_x}}\int_0^\infty \!\!\! \frac{db}{\sqrt{l_y}} \,
\G_R(a,b;\tau_o)\,\exp \left(\frac{i k a^2}{2l_x}+\frac{i k
b^2}{2l_y}\right)\ ,\cr &&
\ea
where we used\equ{G0} in the latter expression. Here $a$ and $b$
denote the radial deviations of $\gamma$ from $\gamma_c$ at the
points $\bx_o$ and $\by_o$ of Fig.~2. We will see in Sec.~4.4 that
the propagation to a point at the surface of the obstacle is only
approximately described by the free Green function. (Roughly
speaking, a photon with large but finite $k$ effectively
penetrates a small distance into the sphere.) The {\it assumption}
that ${\cal G}_r$ separates in the form of\equ{decompGr} neglects
subleading terms of the asymptotic expansion in the penumbra.

$\G_R$ is the coordinate representation of the time evolution
operator generated by the time-independent Hamiltonian
\bl{hamR}
H_R(\pi,x)=\frac{\pi^2}{2 m_E} -\frac{E}{R} x
\ee
in the half-space $x>0$. The integrals over $a$ and $b$
in\equ{decompGr} extend only from zero to infinity because
$\G_R(a,b;\tau)$ vanishes whenever either $a$ or $b$ are negative.
We recognize that the one-dimensional quantum mechanical problem
described by\equ{hamR} is the propagation of a non-relativistic
particle of mass $m_E=E/v^2(E)$ under the influence of a constant
force leading to an acceleration $|g|=v^2(E)/R$ directed {\it
away} from $x=0$. Note that $\G_R(x,y;\tau)= \G^*_R(y,x;-\tau)$,
because  the Hamiltonian\equ{hamR}, although not bounded below, is
hermitian. By definition, the causal Green function  $\G_R$
satisfies the second order differential equation
\bl{diffeq}
-\left(\frac{\hbar^2}{2 m_E} \frac{\partial^2}{\partial x^2} +
\frac{E x}{R}\right)\G_R(x,y;\tau)=i\hbar\frac{\partial}{\partial
\tau} \G_R(x,y;\tau)
\ee
in the half-space $x>0,y>0$ and the initial condition
\bl{inicond}
\G_R(x,y;0)=\delta(x-y)\ .
\ee
The solution to \equ{diffeq} and \equ{inicond} is completely
specified by imposing the boundary conditions
\bl{bound1}
\G_R(0,y;\tau)=0\quad {\rm and}\quad \lim_{x\rightarrow\infty}
\G_R(x,y;\tau)=0\ ,
\ee
and requiring that the magnitude of $\G_R$ remain bounded for
$\tau\rightarrow\infty$
\bl{bounded}
|\G_R(x,y;\tau\rightarrow\infty)|<\infty \ .
\ee
The latter condition excludes any solution which increases
exponentially and is just the requirement that the presence of an
obstacle should not lead to the {\it production} of photons in the
shadowed region.

\subsection{The spectral representation of $\G_{-R}$}
The Hamiltonian\no{hamR} is hermitian and bounded below only for
{\it negative} values of the parameter $R$. In this case it can be
interpreted as describing a non-relativistic particle in a
homogenous gravitational field in the half-space $x>0$ above a
table. The spectral representation of the solution
to~Eqs.\no{diffeq} and\no{inicond} that satisfies the boundary
conditions in~Eqs.\no{bound1} and\no{bounded} is well
known\cite{Sak}. Our strategy is as follows. For $R>0$ waves
diffract and there are no classical orbits, while for $R<0$ there
are classical orbits, a much simpler situation. We will therefore
analyze the $R>0$ case by first considering the $R<0$ case, and
then return to the $R>0$ case of interest by analytic
continuation.

With $R$ replaced by $-R$, we have
\bl{spectral1}
\G_{-R}(x,y;\tau_o)=\frac{k}{\sigma R}
\sum_{n=0}^\infty\Psi_n(x)\Psi_n(y) \exp\left(-i\tau_o
v(E)\epsilon_n\sigma\right)\ ,
\ee
where the inverse length scale $\sigma$ is
\bl{scale}
\sigma=\frac{k}{2^{1/3} (k R)^{2/3}}\ ,
\ee
and the wave function
\bl{wavef}
\Psi_n(x)=\Ai(\frac{k x}{\sigma R}
-\epsilon_n)/|\Ai^\prime(-\epsilon_n)|
\ee
is given by the Airy function\cite{AS} and its derivative,
$\Ai^\prime(z)= (d/dz) \Ai(z)$. Considered as a function on the
complex plane, the Airy function $\Ai(z)$ is the solution to the
differential equation
\bl{Airy}
\frac{d^2}{d z^2} {\rm Ai}(z)=z{\rm Ai}(z)
\ee
with the asymptotic behavior for $|z|\rightarrow\infty$,
\begin{eqnarray}\label{boundAi}
\Ai(z)&\sim& \frac{1}{2\sqrt{\pi}} z^{-1/4}\exp(-\frac{2}{3}
z^{3/2})\ ,\  |{\rm arg} z|<\pi \cr \Ai(-z)&\sim&
\frac{1}{\sqrt{\pi}} z^{-1/4}\sin(\frac{2}{3}
z^{3/2}+\frac{\pi}{4})\ ,\ |{\rm arg} z|<\frac{2\pi}{3}\ .
\end{eqnarray}
The first of the boundary conditions of\equ{bound1} relates the
$\epsilon_n$'s to the zero's of the Airy function
\bl{zeros}
\Ai(-\epsilon_n)=0 \ ,\ \ n=0,1,\dots
\ee
Since the zero's of the Airy function are all on the negative real
axis\cite{AS}, the $\epsilon_n$ are real and positive and are
ordered as $0<\epsilon_n<\epsilon_{n+1}$. That
$\Ai^\prime(-\epsilon_n)$ gives the proper normalization  of the
wave functions follows from the identity,
\bl{norm}
\int_z^\infty dx \Ai^2(x) =\left(\Ai^\prime(z)\right)^2 -z
\Ai^2(z)\ .
\ee
\equ{norm} is proven  by taking the derivative of both sides and
using\equ{Airy}. Choosing $z=-\epsilon_n$ in \equ{norm}, dividing
by $[\Ai^\prime(-\epsilon_n)]^2$, and redefining the integration
variable as $x\rightarrow \sigma x-\epsilon_n$, one finds that
$\int_0^\infty dx\Psi_n^2(x)=1/\sigma$. Since the $\Psi_n$ are
eigenfunctions of a hermitian operator corresponding to the
eigenvalue $\epsilon_n$, $\Psi_n$ and $\Psi_m$ are orthogonal for
$n\neq m$ and therefore,
\bl{normalized}
\int_0^\infty dx \Psi_n(x)\Psi_m(x)=\delta_{nm}/\sigma\ .
\ee
The completeness of the spectrum of the bounded hermitian operator
$H_{-R}$ in the space of normalizable functions on the half space
$x>0$ that vanish at $x=0$, together with\equ{normalized}, proves
that\equ{spectral1} also fulfills \equ{inicond}. Thus the spectral
representation\equ{spectral1} would be the (unique) solution of
\equ{diffeq} and\equ{inicond} we are looking for {\it if} the
parameter $R$ in \equ{diffeq} were negative.

\subsection{The spectral representation of $\G_R$ by analytic continuation}
To obtain the spectral representation of $\G_R$ for {\it positive
real} values of $R$, we analytically continue $\G_{-R}$ in the
complex plane. With $R$ the modulus of the complex number
$R_\phi$,
\bl{analytic}
R_\phi=R e^{i\phi}\ ,
\ee
the analytic continuation of the spectral
representation\equ{spectral1} $\G_{-R_\phi}$ in the phase $\phi$
is uniquely given by requiring that \equ{bounded} hold for all
$0\leq\phi\leq\pi$. The dependence of the inverse scale factor
$\sigma$ on the phase is
\bl{sigma}
\sigma_\phi=\sigma e^{-2i\phi/3}\ .
\ee
For sufficiently large positive values of $x$,
$-\frac{\pi}{3}<{\rm arg }(k x/(\sigma_\phi R_\phi)-\epsilon_n)
\leq 0$ for any fixed $n$ and the Airy function decays
exponentially as given in\equ{boundAi}. Note, however, that the
behavior of the wave functions for large arguments becomes
oscillatory with a power-law decay at the endpoint $\phi=\pi$ of
the interval.  This is consistent with the fact that there is only
one turning point of a trajectory with fixed energy in this case.
The boundary conditions\equ{bound1} thus hold for all
$\phi\in[0,\pi]$. Finally, because
\bl{im}
{\rm Im}\sigma_\phi < 0 \ {\rm for}\ 0<\phi\leq\pi\ ,
\ee
the Green function of\equ{spectral1} {\it decays} for large times
$\tau$ and\equ{bounded} holds. Note that the only dependence of
$\G_R$ on the arc length $l_o$ of the creeping section of the
trajectory is via $\tau=\tau_o=l_o/v(E)$. $\G_R$ thus decays
exponentially with the length $l_o$ of the creeping section with
an exponent that is proportional to $l_o (k R)^{1/3}/R$. This
dependence of the exponent on the cube root of the wave number is
somewhat unusual and perhaps unexpected for a semiclassical
expansion, but is typical for diffraction\cite{Ke62}. Setting
$\phi=\pi$ in\equ{spectral1}, $R\rightarrow R_\pi=-R$ and
therefore $\sigma\rightarrow\sigma_\pi\equiv\bar\sigma$, where
\bl{scalebar}
\bar\sigma=\frac{k e^{-2\pi i/3}}{2^{1/3} (k R)^{2/3}}\
.
\ee
Using~Eqs.\no{spectral1} and\no{wavef}, one thus finally obtains
\bl{GRspec}
\G_R(x,y;\tau_o)=\G_{-R_\pi}(x,y;\tau_o)=\frac{-k}{\bar\sigma R}
\sum_{n=0}^\infty \frac{\Ai(-\frac{k x}{\bar\sigma
R}-\epsilon_n)\Ai(-\frac{k y}{\bar\sigma R}
-\epsilon_n)}{[\Ai^\prime(-\epsilon_n)]^2} \ e^{-i\bar\sigma l_o
\epsilon_n} \ ,
\ee
as the spectral representation of the Green function that
satisfies Eqs.\no{diffeq} and\no{inicond} as well as the boundary
conditions Eqs.\no{bound1} and\no{bounded}.

Inserting\equ{GRspec} in\equ{decompGr}, $\G_r$ is found to have
the representation
\bl{multGr}
\G_r(0,0;\tau_c)=\frac{i k}{4\pi\bar\sigma R}\sum_{n=0}^\infty
{\cal D}_n\left(\bar\sigma l_x\right) {\cal D}_n\left(\bar\sigma
l_y \right)\ e^{-i\bar\sigma l_o\epsilon_n} \ ,
\ee
with amplitudes
\bl{defD}
{\cal D}_n(\xi)=\int_0^\infty
d\rho\,\frac{\Ai(\rho\sqrt{\xi}-\epsilon_n)}{|\Ai^\prime(-\epsilon_n)|}
\exp\left( \frac{i\rho^2}{4}\right)\ ,
\ee
the result previously obtained\cite{KL70} for the asymptotic
behavior in the deep shadow region. Within the semiclassical
approximation it might appear to be consistent to further
approximate the Airy function and its zeros semiclassically and
evaluate the integral in\equ{defD} in saddle point approximation.
This is done in Appendix~B and indeed was part of the original
approach\cite{Ke62,KL70}. However, although perfectly reasonable,
one should perhaps point out that it is not simple to justify this
procedure from the {\it semiclassical} point of view: the error of
the semiclassical approximation is largest for the {\it lowest}
zeros of the Airy function, which apparently dominate the sum
in\equ{multGr} in the semiclassical limit. An alternative to the
semiclassical approximation of the spectral representation of the
Green function and associated integrals is the direct
semi-classical evaluation of the path-integral discussed below. It
gives corrections to the above procedure that become important
outside the deeply shadowed region.

\section{Diffraction by a Sphere: Semiclassical Evaluation of the
Path Integral} In the previous section we constructed the {\it
exact} spectral representation, given by~Eqs.\no{multGr}
and\no{defD}, of the Green-function\equ{Gr} for the radial motion
by analytic continuation. At the end of the calculation we argued
that the semiclassical approximation of the zeros, $\epsilon_n$,
and of the remaining integrals ${\cal D}_n$ may not be fully
justified. We now proceed to do the calculation in reverse; we
first obtain the semiclassical approximation to $\G_r$ for
negative values of the parameter $R$ and {\it then} analytically
continue the semiclassical result to positive values of $R$.
Although this procedure is of only pedagogical value in the case
of a sphere\cite{fnote2}, it has the great advantage of being
generalizable to more complex situations. We will see that this
method of evaluation is also interesting for conceptual reasons.

The point is that there {\it are} classical trajectories for the
radial motion from $x$ to $y$ in time $\tau$ {\it if} the
parameter $R$ in the definition\equ{pot} of the potential
$V_E(x_r;t)$ of the Hamiltonian $H_r$ given in\equ{hamr} is
negative. In this case the force on the particle is either absent
or directed toward the endpoint at $x_r=0$. The force of magnitude
$E/R$ is turned on at ``time'' $\tau_x$ and lasts only for a
finite time $\tau_o$. Using the horizontal axis to denote the time
and the vertical one to denote the ``height'' $x(t)$, the
classical trajectory during the time interval $\tau_o$ is that of
a ball bouncing up and down vertically on a table. See Fig.~3.
There are many classical trajectories that start with an initial
height $x(0)=x$ and which, at $\tau_c=\tau_x+\tau_o+\tau_y$,
attain a final height $x(\tau_c)=y$. They are distinguished by the
number of bounces\cite{fnote3}, $n=0,1,2,\dots$. If the initial or
final point are above the table, the classification of a classical
trajectory by the number of bounces is not unique. There are
classical trajectories with the same number of bounces that differ
in the directions of the initial and/or final momentum. The
ambiguity for a given $n$ is at most 4-fold. If $x\leq 0$ and
$y\leq 0$, this ambiguity is removed since the ball has to be
above the table in the intermediate region. The signs of the
initial and final momenta must be positive and negative,
respectively, and the classification of a classical trajectory by
the number $n$ is unique. Eventually we are interested only in
trajectories that start and end level with the table. To avoid
unnecessary (and irrelevant) complications, we consider only the
situation where the initial and final points satisfy $x\leq 0$ and
$y\leq 0$, that is, are below or level with the table.
{\vskip0.1truecm\epsfig{figure=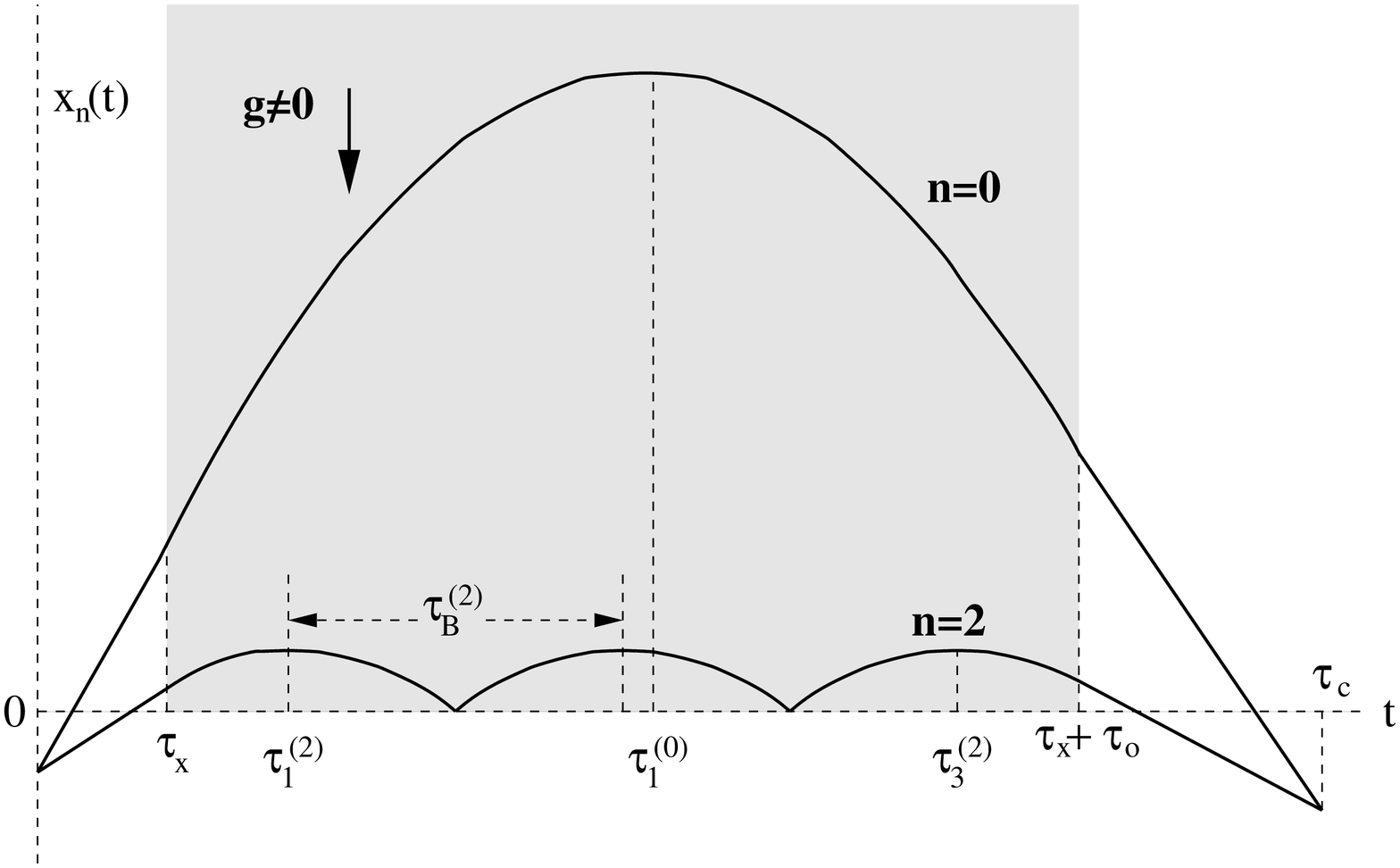,height=5.0truecm}\nobreak

{\small\noindent Fig.~3: Radial deviations with $n=0$ (no bounces)
and $n=2$ (two bounces) from a particular classical path
$\gamma_c$ for a spherical obstacle of radius $R$. In the time
intervals $[0,\tau_x]$ and $[\tau_x+\tau_o,\tau_c]$, there is no
acceleration and these sections of the paths are straight lines;
the motion in the interval $[\tau_x,\tau_x+\tau_o]$ is governed by
a constant acceleration $g=-v^2(E)/R$. The initial and final
radial deviations from $\gamma_c$, $x=x_n(0)$ and $y=x_n(\tau_c)$,
not labeled in the figure, do not depend on $n$. $\tau_i^{(n)}$ is
the $i$-th turning point of the path with $n$ bounces,
$\tau_B^{(n)}$ is the time between successive bounces, or
successive turning points.}

The classical trajectory $\gamma_n$ connecting $x$ and $y$ in time
$\tau_c$ with $n$ bounces is a piecewise connected solution to the
equation of motion
\bl{em}
{\stackrel{..}\atop{\!\!\!\!\! x_n}}(t)=-g(t)
\ee
with $x(0)=x$ and $x(\tau_c)=y$ and $x(t)\geq 0$ for
$\tau_x<t<\tau_x+\tau_o$ that apart from the $n$ points
corresponding to bounces is continuous in phase space. Since the
(negative) acceleration $g(t)=v^2(E)/R=g$ does not depend on time
in the interval $\tau_x<t<\tau_x+\tau_o$, the ``energy'' (for
notational convenience we extract the scale to have a
dimensionless energy)
\bl{ener}
e_n=\frac{2}{m_E g R}\left(\frac{\pi_n^2(t)}{2 m_E} + m_E g
x_n(t)\right)= \frac{\pi_n^2(t)}{(\hbar k)^2} +2\frac{x_n(t)}{R}\
,
\ee
is a real positive constant of motion that characterizes the
trajectory in this time interval. In the initial and final time
intervals $t\in [0,\tau_x]$ and $t\in[\tau_x+\tau_o,\tau_c]$ the
acceleration vanishes and the momenta are constant. We also
introduce dimensionless initial and final ``momenta'',
\begin{eqnarray}\label{mom}
p_n&=&~\pi_n(0)/(\hbar k)=~ \pi_n(\tau_x)/(\hbar k)>0\ ,\cr
p_n^\prime &=&-\pi_n(\tau_c)/(\hbar
k)=-\pi_n(\tau_x+\tau_o)/(\hbar k)>0\ .
\end{eqnarray}
The signs have been chosen so that $p_n$ and $p_n^\prime$ are both
positive for  initial and final points $x\leq 0$ and $y\leq 0$.
The equation of motion\no{em} with $g(t)=0$ implies that
\bl{heights}
x_n(\tau_x)=x +l_x p_n \ \ {\rm and}\ \ x_n(\tau_x+\tau_o) =y+l_y
p_n^\prime\ .
\ee
Demanding continuity of the classical trajectory in phase space at
$t=\tau_x$ and $t=\tau_x+\tau_o$, we determine $p_n$ and
$p_n^\prime$ in terms of $e_n$. They are given by
\begin{eqnarray}\label{qs}
p_n&=&\sqrt{e_n+(l_x/R)^2-2 x/R} -l_x/R\cr
p_n^\prime&=&\sqrt{e_n+(l_y/R)^2-2 y/R}-l_y/R\ .
\end{eqnarray}
The signs of the square roots have been chosen so that both scaled
momenta, $p_n$ and $p^\prime_n$,  are positive for initial and
final heights $x\leq 0$ and $y\leq 0$.

To completely specify the trajectory, we need an equation for
$e_n$. The acceleration is constant for $\tau_x<t<\tau_x+\tau_o$
and the first turning point of the trajectory occurs at
$t=\tau_1^{(n)}$, where
\bl{turn1}
\tau_1^{(n)}=\tau_x+ p_n R/v(E)\ .
\ee
The last turning point, the $(n+1)^{\rm th}$, occurs at
$t=\tau_{n+1}^{(n)}$, where
\bl{turnn}
\tau_{n+1}^{(n)}=\tau_x+\tau_o-p_n^\prime R/v(E)\ .
\ee
The equation of motion\no{em} for constant acceleration finally
relates the time $\tau_B^{(n)}$ between successive bounces (or
successive turning points) when there are $n$ bounces in all to
the conserved "energy" $e_n$,
\bl{bouncetau}
\tau_B^{(n)}=2 \sqrt{e_n} R/v(E)\ .
\ee
The time  to the first turning point plus the time for $n$ bounces
plus the time from the last turning point to the end of the
trajectory is the total time $\tau_c$ of the trajectory. We thus
have that $\tau_1^{(n)}+n \tau_B^{(n)}+
(\tau_c-\tau_{n+1}^{(n)})=\tau_c$, or
\bl{ensol}
p_n+p_n^\prime+ 2 n \sqrt{e_n}=l_o/R\ .
\ee
\equ{ensol} together with \equ{qs} is a fourth order algebraic
equation for $e_n$ whose roots can be found analytically. There is
only one real and positive solution $e_n$, because the left hand
side of\equ{ensol} is a monotonically increasing function of $e_n$
that vanishes at $e_n=0$ and goes as $2 (n+1) \sqrt{e_n}$ for
large values of $e_n$. We remark that  $e_n,p_n$ and $p_n^\prime$
given by\equ{ensol} and\equ{qs} depend only on the lengths
$l_x,l_o$ and $l_y$ of the problem and the scaled initial and
final heights $x$ and $y$ (all measured in units of $R$). The
solution $e_n$ of\equ{ensol} therefore does not depend on $E$ nor
$v(E)$. The action $S_n(x,y)$ of the classical trajectory with $n$
bounces is proportional to $E R/v(E)=\hbar k R$. After some
algebra, $S_n(x,y)$ can be expressed in terms of $e_n,p_n$ and
$p_n^\prime$ implicitly given by \equ{ensol} and \equ{qs}. We find
\bl{sn}
\frac{S_n(x,y)}{\hbar k R}=\frac{p_n^3+p_n^{\prime\,3}+2 n
e_n^{3/2}}{3} + \frac{p_n^2 l_x+p_n^{\prime\,2} l_y -e_n l_o}{2R}\
.
\ee
Using\equ{qs}, one explicitly verifies that
\bl{statione}
\frac{\partial S_n(x,y)}{\partial e_n}=\frac{\hbar k
R}{2}(p_n+p_n^\prime+ 2 n \sqrt{e_n}-l_o/R)\ .
\ee
\equ{ensol} thus selects the energy $e_n$ at which the action is
stationary. This fact will be of some consequence when we compare
the present approach to the one of section~3.

The semiclassical approximation $G_r(x,y;\tau_c,-R)$ to the Green
function\break $\G_r(x,y;\tau_c,-R)$ defined in\equ{Gr} might {\it
seem} to be the sum of the asymptotic contributions from the
infinite number of classical path. The semiclassical contribution
of a single classical path is known; it is given by Van Vleck's
formula\cite{VF}.  {\it If} we could simply sum over the
asymptotic contributions of all classical trajectories, we would
(in this one dimensional case) obtain
\bl{semir1}
G_r(0,0;\tau_c,-R)=\frac{1}{\sqrt{2\pi
i\hbar}}\sum_{n=0}^\infty\left.\sqrt{\frac{-\partial^2
S_n(x,y)}{\partial x\partial y}}\right|_{x=y=0}\!\!\!\! \exp i
\left(\frac{S_n(0,0)}{\hbar} -\frac{3\pi n}{2}\right).
\ee
(There are $n$ conjugate points and $n$ points of reflection on
the classical trajectory with $n$ bounces, contributing in total a
phase factor of $\exp(-i 3\pi n/2)$.)

Although each term of the sum is the correct asymptotic
contribution from a {\it particular} classical trajectory, we now
show that\equ{semir1} does {\it not} give the asymptotic behavior
of the Green function $\G_r$ for $kR\rightarrow\infty$. The cause
of the problem is that for any fixed value of $kR$, the classical
trajectories and their corresponding action $S_n$ become
indistinguishable as $n\rightarrow\infty$. The "floor" at $x=0$ is
a zero-dimensional caustic of the classical motion\cite{lit}. It
then is not possible to interchange the implicit
$kR\rightarrow\infty$ limit with the sum over $n$ in\equ{semir1}.

To see that the interchange is not allowed, we examine the
asymptotic behavior of the classical action $S_n$ for trajectories
with large $n$, assuming for the moment that\equ{semir1} is valid.
We begin by observing that the second derivatives of the action
appearing in\equ{semir1} can be evaluated in terms of the
solutions $p_n,p^\prime_n$ and $e_n$ of \equ{ensol} and \equ{qs}
for $x=y=0$. The classical equations of motion imply that the
initial momentum $\pi_n(0)$ is given by
\bl{classem}
\frac{\partial S_n(x,y)}{\partial x}=-\pi_n(0)=-\hbar k p_n .
\ee
\equ{classem} can be checked explicitly by taking the derivative
of\equ{sn} and using~Eqs.\no{qs} and\no{ensol}. From\equ{classem}
we thus have that
\bl{ddS}
-\left.\frac{\partial^2 S_n(x,y)}{\partial x\partial
y}\right|_{x=y=0}= \left.\hbar k \frac{\partial p_n}{\partial
y}\right|_{x=y=0}\ .
\ee
The variation of the initial momentum with the endpoint of the
trajectory required in the latter expression is implicitly given
by\equ{ensol} and the definitions\equ{qs}. After some algebra we
obtain
\bl{ddSexp}
-\frac{\partial^2 S_n(x,y)}{\partial x\partial y}=\frac{\hbar k
e_n}{ l_o (p_n+\frac{l_x}{R})(p_n^\prime+\frac{l_y}{R}) +p_n l_x
(p^\prime_n+\frac{l_y}{R})+p^\prime_n l_y (p_n+\frac{l_x}{R})}\ .
\ee
Note that in\equ{semir1} we need only know the right hand side of
\equ{ddSexp} at $x=y=0$. To evaluate the function\equ{semir1} it
thus is sufficient to solve\equ{ensol} with $x=y=0$.

The solution $e_n$ of\equ{ensol} will be small compared to either
$l_x^2/R^2$ or $l_y^2/R^2$ when $n$ is sufficiently large.  For
$x=y=0$, we then may expand the roots occurring in\equ{qs} and
thus solve \equ{ensol} asymptotically for large $n$ with the
result
\bl{enas}
\sqrt{e_n}=\frac{1}{2 n} \left(l_o/R\right)
 +O(n^{-3},x/l_x, y/l_y)\ .
\ee
The scaled momenta $p_n$ and $p^\prime_n$ defined by\equ{qs}
therefore are of order $n^{-2}$ and
\bl{ddSlargen}
\left.-\frac{\partial^2 S_n(x,y)}{\partial x\partial
y}\right|_{x=y=0}=\frac{\hbar k l_o}{ 4 l_x l_y n^2}+O(n^{-4})\ .
\ee
Note that in this limit
\bl{Sn0}
S_n(0,0)=-\frac{\hbar k l_o^3}{24 R^2 n^2} +O(n^{-4})
\ee
depends neither on $l_x$ nor on $l_y$. This is readily understood.
For $n$ bounces within a given time
 interval  $\tau_c$, their height becomes exceedingly small compared
 to $l_x$ or $l_y$. For $x=y=0$ and large $n$, the initial and
 final momenta of the trajectory are of order $n^{-2}$ and the
 free sections of the trajectory  therefore do not contribute
 appreciably to the classical action. One can check that \equ{Sn0}
 indeed is just the action
 for a classical trajectory with $n$ bounces that
 altogether take a time  $\tau_o=l_o/v(E)$. The asymptotic
 behavior in~Eqs.\no{ddSlargen} and\no{Sn0} implies that the sum
 in\equ{semir1} converges to a logarithm for large $n$.

The asymptotic form\equ{Sn0}, however, also shows that the actions
of classical trajectories differ by much less than $\hbar$ for
$n\gg \sqrt{k R}$. For a fixed value of $kR$, the classical paths
with sufficiently large $n$ therefore could (and should) be
considered quantum fluctuations of each other. It becomes a matter
of exchanging limits: the asymptotic weight of a particular
classical path characterized by $n$ bounces in the limit $k
R\rightarrow \infty$ is indeed given by its contribution
to\equ{semir1}. However, we actually would like to know how the
(infinite) sum of classical paths contribute to the Green function
for a large, but fixed value of $k R$. The problem arises because
the trajectory $\lim_{n->\infty} x_n(t)= x_\infty(t)=0$ is a
caustic that is approached by an infinite number of classical
trajectories. Note, however, that the sum in\equ{semir1} would be
cut off at a finite value $n=n_{max}$ for a {\it negative} value
of the initial point $x$ or final point $y$. Due to the
non-vanishing initial and/or final momentum, the ``energy'' $e_n$
of a trajectory in this case is bounded from below, $e_n\ge {\rm
Max}\left((x/l_x)^2, (y/l_y)^2\right)$, and \equ{ensol} can be
solved only for $n\le n_{\rm max}$. \equ{semir1} in this case {\it
would} give the correct asymptotic expression for sufficiently
large $kR$.

For a given geometry and wave number $k$,  successive terms of the
sum in\equ{semir1} tend to interfere destructively for
sufficiently high $n$ -- one thus should reorder the sum and
collect terms of similar phase. To do so, we make use of our
earlier observation in\equ{statione} that\equ{ensol} gives the
stationary point of $S_n$ with respect to a variation of the
``energy'' $e_n$. The second derivative of the action at this
stationary point is
\bal{der2e}
\frac{\partial^2 S_n(0,0)}{\partial e_n^2}&=&\frac{\hbar kR}{4}
\left.\left(\frac{1}{p_n+l_x/R}+\frac{1}{p^\prime_n+l_y/R}-
\frac{p_n+p^\prime_n-l_o/R }{e_n}\right)\right|_{x=y=0}\cr
&&\hskip-5em =\left.\frac{\hbar k [l_o
(p_n+\frac{l_x}{R})(p^\prime_n+\frac{l_y}{R})+p_n l_x
(p^\prime_n+\frac{l_y}{R})+p^\prime_n l_y (p_n+\frac{l_x}{R})] }{4
e_n (p_n+l_x/R)(p^\prime_n+l_y/R)}\right|_{x=y=0}>0\cr &&\ ,
\ea
where we have used the definitions\no{qs} to write the numerator
in terms of $p_n$ and $p_n^\prime$ only.

We will show just below that it is consistent, within the
semiclassical approximation, to replace \equ{semir1} by the
integral expression
\bl{intsemir1}
G_r(0,0;\tau_c,-R)=\frac{-i}{\pi \hbar}\int_0^{\infty} e\,de\,
B(e) e^{i S(e)/\hbar} \sum_{n=0}^\infty e^{i n f(e)}\ ,
\ee
where
\bl{fdef}
f(e)= \frac{2 k R}{3}e^3 -\frac{3\pi}{2}\ .
\ee
 In\equ{intsemir1} the factor $B(e)$ is given by
Eqs.\no{ddSexp} and\no{der2e} with $e_n$ replaced by $e^2$, that
is,
\bal{De}
B(e)&=&\left.\sqrt{\frac{-\partial^2 S_n(x,y)}{\partial x\partial
y}\frac{\partial^2 S_n(x,y)}{\partial
e_n^2}}\right|_{x=y=0;e_n=e^2}\cr & =& \frac{\hbar k}{2
\left[(e^2+(l_x/R)^2)(e^2+(l_y/R)^2)\right]^{1/4}}\ ,
\ea
and $S(e)$ is related to the action of\equ{sn},
\bl{Se}
S(e)=\hbar kR\left(\frac{p^3+p^{\prime\,3}}{3} + \frac{p^2
l_x+p^{\prime\,2} l_y -e^2 l_o}{2R}\right)\ ,
\ee
where here too $e_n$ has been replaced by $e^2$. In addition the
only explicitly $n$-dependent term in $S_n(x,y)$ of\equ{sn} has
been separated out. In~Eqs.\no{De} and\no{Se} the variables $p$
and $p^\prime$ are the dimensionless initial and final momenta
of\equ{qs} for $x=y=0$ and "energy" $e^2$, that is
\bl{qe} p=\sqrt{e^2+(l_x/R)^2} -l_x/R\  ,\ \ p^\prime=\sqrt{e^2+(l_y/R)^2}-l_y/R\ .
\ee

To show the semiclassical equivalence of~Eqs.\no{semir1}
and\no{intsemir1}, recall that the phase of the integrand
in\equ{intsemir1} for a given $n$ is stationary at $e=\sqrt{e_n}$.
Expanding about $e=\sqrt{e_n}$, and evaluating the resultant
Gaussian integral, one finds that~Eqs.\no{semir1}
and\no{intsemir1} are indeed equivalent term by term.

The asymptotic contribution of a classical path with a given $n$
to the Green function in\equ{intsemir1} thus is precisely the same
as in\equ{semir1}. On the other hand we now can interchange the
limits and obtain the correct asymptotic expansion by summing the
geometrical series in\equ{intsemir1} for a given value of $kR$.
With a small positive imaginary component of $e$, the sum
in\equ{intsemir1} is evaluated as
\bl{sum}
\lim_{\eta\rightarrow 0^+}\sum_{n=0}^\infty e^{in
(f(e)+i\eta)}=\lim_{\eta\rightarrow 0^+} \frac{1}{1-e^{i
(f(e)+i\eta)}} .
\ee

The integrand of\equ{intsemir1} therefore has simple poles just
below the real axis at $f(e)=2\pi m-i\eta$ for integer $m$. Apart
from these simple poles, the integrand is analytic in the shaded
region of Fig.~4. The integration over $e$ in\equ{intsemir1} along
the real axis therefore is equivalent to integrating along the
contours ${\cal C}_{sd}$ and ${\cal C}_{\cal R}$ shown in Fig.~4
and including the residues from the enclosed poles. The paths
${\cal C}_{sd}$ and ${\cal C}_{\cal R}$ in the complex plane are
chosen so that, as shown below,
\begin{itemize}
  \item [i)] the contribution from ${\cal C}_{\cal R}$ to the
  integral is negligible for sufficiently large ${\cal R}$, and
  \item[ii)] the modulus of the integrand in\equ{intsemir1}
  decreases monotonically and as fast as possible along ${\cal C}_{sd}$,
  the path of steepest descent.
\end{itemize}
The semi-classical evaluation of the integral in\equ{intsemir1}
thus has three distinct contributions, to be denoted by $G_{\rm
poles},\ G_{\cal C_R}$ and $G_{{\cal C}_{\rm sd}}$. We discuss
them separately.
{\vskip0.1truecm\epsfig{figure=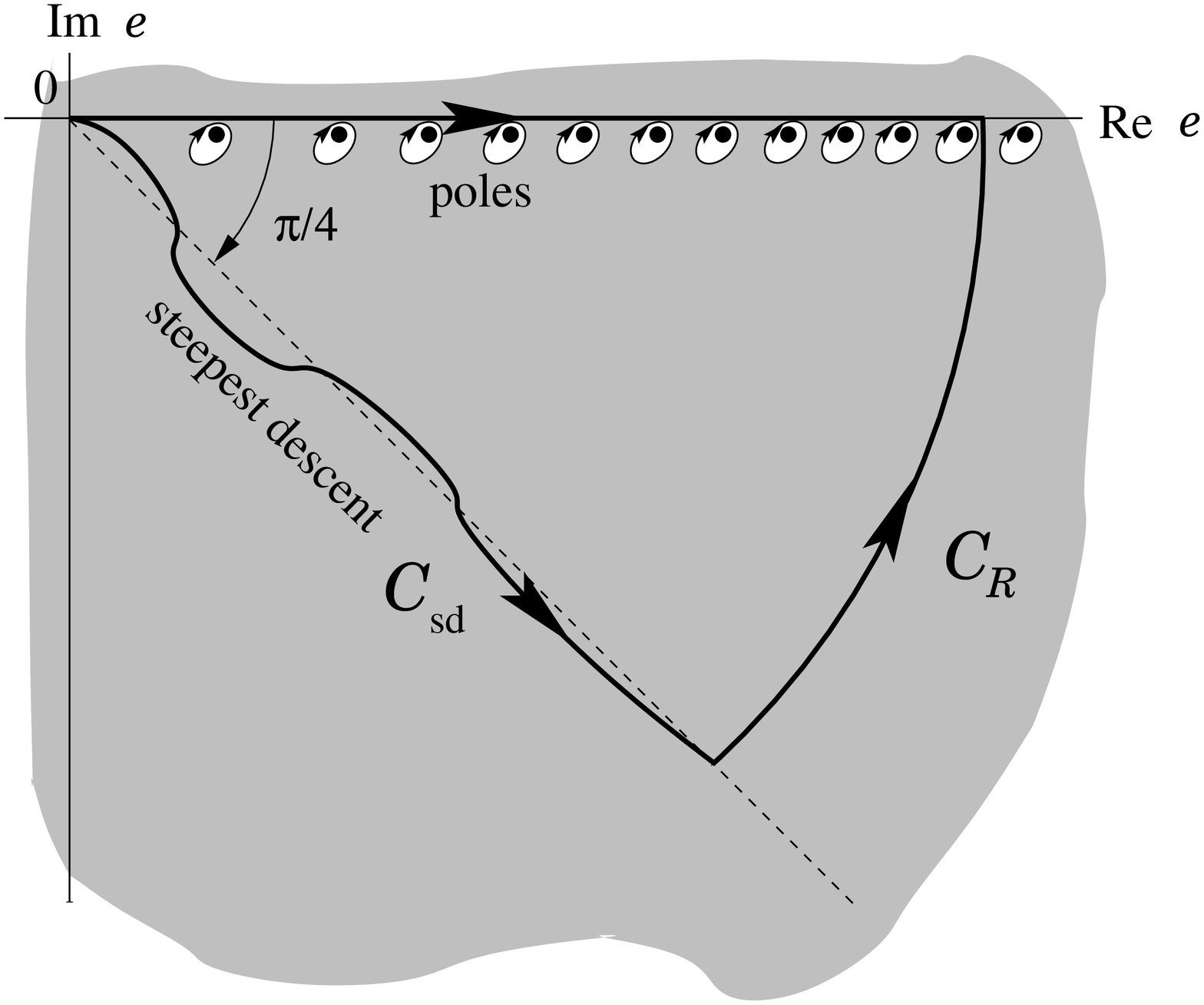,height=6.0truecm}\nobreak

{\small\noindent Fig.~4: Contours of integration. Apart from
discrete poles just below the positive real axis,  the integrand
of\equ{intsemir1} is analytic in the shaded region of the complex
plane. The integral along the positive real axis in\equ{intsemir1}
from the origin to ${\cal R}$ is evaluated by integrating along
the contours ${\cal C}_{\rm sd}$ and ${\cal C}_{\cal R}$, taking
the residues of the poles inside the closed contour into account.
${\cal C}_{\rm sd}$ is the path of steepest descent of the
integrand that starts at the origin. It asymptotically approaches
the dashed line with arg~$e= -\pi/4$ for large $|e|\sim\infty$.}

\subsection{The Contribution from Poles with Integer $m\ge 0$}
In the shaded region of the complex plane of Fig.~4, the integrand
of\equ{intsemir1} is analytic except for simple poles at
\bl{evs}
e=e(m)=\left(\frac{3\pi(m+3/4)}{k
R}\right)^{1/3}=\left(\frac{2}{kR}\right)^{1/3}\sqrt{\bar\epsilon_m}\
,
\ee
just below the real axis for any integer $m\geq 0$. Note that
$\bar\epsilon_m$ corresponds to a semiclassical determination of
the $(m+1)^{\rm th}$ zero $\epsilon_m$ of the Airy
function\cite{Sak}. The contribution $G_{\rm poles}(-R)$ to
$G_r(0,0;\tau_c,-R)$ from the poles is
\bl{resid} -2\pi i\sum_{m=0}^\infty
\left\{\begin{array}{c}\textrm{residue of}\cr \textrm{pole
at~}e(m)\end{array}\right\} =\frac{-i}{\hbar k R
}\sum_{m=0}^\infty \frac{ B(e(m))}{e(m)} e^{i S(e(m))/\hbar}\ .
\ee
Using Eqs.\no{scale}\no{De},\no{qe},\no{Se} and\no{evs}, $G_{\rm
poles}(-R)$ can be rewritten as
\bl{semsing}
G_{\rm poles}(-R)=\frac{k}{4\pi i\sigma R}\sum_{n=0}^\infty
D_n\left(\sigma l_x\right) D_n\left(\sigma l_y\right) e^{-i\sigma
l_o\bar\epsilon_n}\ ,
\ee
with diffraction amplitudes
\bl{Dsemi}
D_n(\xi)=\frac{\sqrt{\pi}}{[\bar\epsilon_n(\bar\epsilon_n+\xi^2)]^{1/4}}
\exp i\left(\frac{2\bar p_n^3(\xi)}{3}+\xi \bar p_n^2(\xi)\right)\
.
\ee
The $D_n(\xi)$ depend on $\bar\epsilon_n$ defined by\equ{evs} and
the corresponding (rescaled) "momenta"
\bl{barq}
\bar p_n(\xi)=\sqrt{\bar\epsilon_n+\xi^2}-\xi\ .
\ee

Since $\bar p_n(\xi)\sim e(n)\sim n^{1/3}$ for large $n$, the
$n^{\rm th}$ and $(n+1)^{\rm th}$ contribution to the sum
in\equ{semsing} differ by a {\it finite} phase. By first summing
in\equ{intsemir1} rather than semiclassically evaluating the
integral over $e$ for each summand, one avoids the problem
discussed earlier that for fixed wave number $k$, classical paths
near a caustic interfere with one another. [Note that the
individual summands of\equ{semsing} do not correspond to
contributions from individual classical paths of the original
problem.]

The similarity between the expressions in Eqs.\no{semsing}
and\no{multGr} is no coincidence. We show in Appendix~B that apart
from approximating the zero's of the Airy function $\epsilon_n$ by
$\bar\epsilon_n$, $D_n$ is proportional to the integral
of\equ{defD} evaluated in saddle point approximation. We thus
could have obtained the expression\equ{semsing} by evaluating all
integrals in saddle-point approximation and replacing the
wave-functions $\Psi_n(x)$ and corresponding energies $\epsilon_n$
in\equ{multGr} by their WKB counterparts. The fact that such a
procedure ignores any asymptotic contribution from the integration
along ${\cal C}_{sd}$ and ${\cal C}_{\cal R}$ indicates that this
may not be the whole story, and indeed it is not.

\subsection{The Contribution $G_{\cal C_R}$}
The contribution to the Green function from integrating along the
arc ${\cal C}_{\cal R}$ is negligible for sufficiently large
${\cal R}$. To see this, we use~Eqs.\no{intsemir1}-\no{sum} to
write this contribution to $G_r$ in the form
\bl{integrand2}
G_{\cal C_R}(-R)= \frac{i}{\pi \hbar}\int_{{\cal C}_{\cal R}}
e\,de\, B(e) \frac{e^{i [(S(e)/\hbar) -f(e)]}}{1-e^{-i f(e)}}\ ,
\ee
where the contour ${\cal C}_{\cal R}$ is shown in Fig.~4.

Consider the behavior of the integrand in\equ{integrand2}, with
the contour ${\cal C}_{\cal R}$ chosen to be an arc of very large
radius ${\cal R}=|e|$ centered on the origin that extends from the
positive real axis into the lower half of the complex plane with
$-\pi/3<{\rm arg}~e<0$. The real part of $if(e)$ on the arc is
positive and proportional to $|e|^3$. $e^{-i f(e)}$ therefore is
exponentially small on ${\cal C}_{\cal R}$ and for sufficiently
large ${\cal R}=|e|$ is negligible compared to $1$ -- the
denominator of the integrand thus approaches unity. To expand
$(S(e)/\hbar)- f(e)$ for large $|e|$ and $-\pi/3<{\rm arg}~e<0$,
note that $p\sim e-l_x$ and $p^\prime\sim e-l_y$ in this region.
The leading $e^3$ term of $S(e)$ for $|e|\sim\infty$ is cancelled
by the leading term of $f(e)$ and one has, for $|e|\sim\infty$ and
$-\frac{\pi}{2}<{\rm arg}~e<0$,
\bl{expSf}
(S(e)/\hbar)-f(e)\sim -\frac{k}{2}(l_x+l_y+l_o)e^2,\ \ .
\ee
Since ${\rm Re}(-i e^2)<0$, \equ{expSf} shows that the
contribution of ${\cal C}_{\cal R}$ to the integral
in\equ{integrand2} vanishes in the limit ${\cal
R}\rightarrow\infty$.

\subsection{The Contribution $G_{{\cal C}_{\rm sd}}$}
The contribution $G_{{\cal C}_{\rm sd}}(-R)$ is given
by\equ{integrand2}, with ${\cal C_R}$ replaced by ${\cal C}_{\rm
sd}$. The section ${\cal C}_{\rm sd}$ of the contour is chosen to
coincide with the path of steepest descent that begins at the
origin. From\equ{expSf} we see that ${\rm arg}~e$ along this path
approaches $-\pi/4$ for very large values of $|e|$. Because $p$
and $p^\prime$ are both proportional to $e^2$ for small $|e|$, the
leading behavior in this limit is (using Eqs.\no{fdef} and\no{Se})
\bal{expSsmall}
(S(e)/\hbar)- f(e) &\sim& \frac{3\pi}{2} -\left[\frac{kl_o}{2}
e^2+\frac{2kR}{3} e^3\right]+ O(e^4), \cr &&\quad
-\frac{\pi}{2}<{\rm arg}~e<0\ .
\ea

When $l_o>0$, the path ${\cal C}_{\rm sd}$ of steepest descent
starts at the origin of the complex plane with ${\rm
arg}~e=-\pi/4$. Unless $R$ is very large compared to $l_x$ and
$l_y$, the terms of order $e^4$ of\equ{expSsmall} are negligible
in the semiclassical approximation. However, at grazing angles,
$l_o/R\sim 0$, the leading dependence on the wave number comes
from the term of order $e^3$ in\equ{expSsmall}. To obtain a
semiclassical expansion that is uniformly  valid in the whole
classically shadowed region, one therefore has to retain all
exponential terms with exponents that are of cubic or lower order
for small $e$. To semiclassical accuracy we have,
using\equ{expSsmall}, $\sigma$ given by\equ{scale}, $B(0)$
by\equ{De} and $f(e)$ by\equ{fdef},
\bal{steep}
G_{{\cal C}_{\rm sd}}(-R)&=&\frac{i}{\pi \hbar}\int_{{\cal C}_{\rm
sd}} e\,de\, B(e) \frac{e^{i [(S(e)/\hbar) -f(e)]}}{1-e^{-i
f(e)}}\cr &\sim& \frac{B(0)}{2\pi i \hbar
(kR)^{2/3}}\,P\left(\left(\frac{3}{4}\right)^\frac{2}{3}\sigma
l_o\right)= \frac{(kR)^{1/3} }{4 \pi i \sqrt{l_x l_y}}\,
P\left(\left(\frac{3}{4}\right)^\frac{2}{3}\sigma l_o\right).\
\ea
(Note that this contribution to the Green function neatly
separates into the product of free one-dimensional Green functions
over distances $l_x$ and $l_y$ and a factor that depends on $l_o$
and $R$.)  For positive real values of $z$, the function $P(z)$
has the integral representation,
\bl{defP}
P(z)=i\sqrt[3]{18}\int_{{\cal C}_{sd}}  \xi d\xi  \, \frac{e^{-i
(z \xi^2+\xi^3)}}{1+i e^{-i\xi^3}}\ .
\ee
$P(z)$ is an entire function that is defined in the whole complex
plane by the convergent series (derived in Appendix~C),
\bl{seriesP}
P(z)=i\left(\frac{2i}{3}\right)^{1/3}\sum_{n=1}^\infty \left(z
e^{-5\pi i/6}\right)^{n-1} \frac{\Gamma(2n/3)}{\Gamma(n)}\, {\rm
Li}_{2n/3}(-i)\ ,
\ee
where
\bl{Li}
{\rm Li}_\nu(\rho)= \sum_{n=1}^\infty \frac{\rho^n}{n^\nu}
\ee
is the poly-logarithm of order $\nu$. Of particular interest are
the asymptotic values of $P(z)$ for large and small $z$,
corresponding to large and small values of $k l_o$:
\bal{asP}
P(0)&=&i\left(\frac{2i}{3}\right)^{1/3} \Gamma(2/3)\, {\rm
Li}_{2/3}(-i)\approx 0.966 e^{0.00347\pi i}\cr P(z)&\sim &
\frac{3^{2/3}e^{-\frac{\pi i}{4}}}{2^{7/6} z}\approx \frac{0.927
e^{-\frac{\pi i}{4}}}{z},\ {\rm for}\ |z|\sim\infty,\
-\frac{\pi}{2}<{\rm arg}~z<\frac{\pi}{2}\ .\cr &&
\ea
The asymptotic forms of\equ{asP} are most easily obtained
from\equ{defP}; see Appendix~C.

\cut{Using the expansion\no{seriesP}, $G_{{\cal C}_{\rm sd}}$
follows from\eq may written in a form similar to that of $G_{\rm
poles}$, that is,
\bl{GrminusR}
G_{{\cal C}_{\rm sd}}=\frac{k}{4\pi i\sigma R}\sum_{n=0}^\infty
\bar D_n\left(\sigma l_x\right)
\bar D_n\left(\sigma l_y\right)\,
\left[\left(\frac{3}{4}\right)^\frac{2}{3}\sigma l_o e^{-5\pi
i/6}\right]^n\ ,
\ee
with $\sigma$ given by\equ{scale} and  the amplitudes $\bar D_n$
defined by,
\bl{barD}
\bar D_n(\xi)=\frac{e^{\pi i/3}}{6^{1/6}}
\sqrt{\frac{\Gamma(2(n+1)/3){\rm Li}_{2(n+1)/3}(-i)}{\xi
\Gamma(n+1)}}\ .
\ee
While the contribution of\equ{semsing} from poles depends
exponentially on $k l_o/(kR)^{2/3}$, the integration along ${\cal
C}_{sd}$ gives terms that are proportional to powers of
$kl_o/(kR)^{2/3}$. The latter are important in the penumbra.}

To finally obtain $G_r(0,0;\tau_c,R)$ we again analytically
continue the expressions in Eqs.\no{steep} and\no{semsing} to
negative values of $R$ in such a manner that $G_r$ remains bounded
for large values of $l_o$. We thus find that the semiclassical
approximation to the Green function for the radial coordinate from
a single extremal path whose end-point is in the shadow of a
sphere of radius $R$ is
\bl{GRfinal}
G_r(0,0;\tau_c,R)=\frac{i k}{4\pi
\bar\sigma R}\left[\frac{P\left((3/4)^\frac{2}{3}\,\bar\sigma l_o\right)
}{2^{2/3}\bar\sigma\sqrt{l_x l_y}}+\sum_{n=0}^\infty
D_n\left(\bar\sigma l_x\right) D_n\left(\bar\sigma l_y\right)
e^{-i\bar\sigma l_o\bar\epsilon_n}\right]
\ee
The diffraction amplitudes $D_n(\xi)$ are defined by
Eqs.\no{Dsemi} and\no{barq}, with $\bar\epsilon_n$ given
by\equ{evs} and $\bar\sigma$ by\equ{scalebar}. The function $P(z)$
for complex $z$ is defined by the expansion\equ{seriesP}.

\subsection{The Penumbra near the Glancing Ray with $l_o=0$}
The correction from the contour ${\cal C}_{\rm sd}$
in\equ{GRfinal} is essential for describing the penumbra. We
compare our result with the asymptotic form of the exact solution
in the penumbra obtained in\cite{Pr96,const}. In the penumbra
$kl_o\lless kR$ and one can expand $G_r$ in powers of $kl_o$. The
leading term in this expansion of $G_r$ is
\bl{expandlo}
\left. G_r(0,0;\tau_c,R)\right|_{l_o=0}=\frac{i k}{4\pi
\bar\sigma R}\left[\frac{P(0)}{2^{2/3}\bar\sigma\sqrt{l_x l_y}}
+\sum_{n=0}^\infty D_n\left(\bar\sigma l_x\right)
D_n\left(\bar\sigma l_y\right)\right] \ .
\ee
The sum in\equ{expandlo} is exponentially damped and converges for
any value of $kR$.  Let $\xi=\bar\sigma l$, with $l=l_x$ or $l_y$;
$\bar\sigma$ is given by\equ{scalebar}.  For $\bar\epsilon_n\gg
|\xi|^2$, we find on using\equ{barq} that
\bl{pexp1}
\bar p_n(\xi)\sim \bar\epsilon_n^{1/2}-\xi,
\ee
and therefore that
\bl{aexp1}
\frac{2}{3}\bar p_n^3(\xi)+\xi\bar p_n^2(\xi)\sim
\frac{2}{3}\bar\epsilon_n^{3/2}-\xi\bar\epsilon_n\ .
\ee
Since ${\rm Im}(\xi)<0$, $D_n(\xi)$ defined by\equ{Dsemi}
therefore decays exponentially and the contribution to the sum
in\equ{expandlo} from terms with $\bar\epsilon_n\gg |\xi|^2$ is
negligible.

We now assume that $\bar\epsilon_n\lless |\xi|^2$. For
$|\bar\epsilon_n/\xi^2|\lless 1$,
\bl{aspn}
\bar p_n(\xi)\sim \bar\epsilon_n/(2\xi) + O(|\xi|^{-3})\ .
\ee
Since $|\xi|\propto (kR)^{1/3}$, we can simplify the sum
in\equ{expandlo} in the asymptotic limit we are interested in.
Retaining only leading terms in $kR$, the definition\equ{Dsemi} of
the diffraction amplitudes leads to the simplification
\bal{exsum1}
\Sigma_D&\equiv&\sum_{n=0}^\infty D_n\left(\bar\sigma l_x\right)
D_n\left(\bar\sigma l_y\right)\cr &\sim
&\frac{\pi}{\bar\sigma\sqrt{l_x l_y}}\sum_{n=0}^\infty
(\bar\epsilon_n)^{-\frac{1}{2}}
\exp\left[i\frac{\bar\epsilon_n^2}{4\bar\sigma} \frac{l_x+l_y}{l_x
l_y}\right]\equiv\frac{\pi}{\bar\sigma\sqrt{l_x
l_y}}\sum_{n=0}^\infty \Phi(n)
\ea
The sum over $n$ in\equ{exsum1} is effectively cut off when
$\epsilon_n^2$ becomes of order $1/\Lambda$ or larger, where
\bl{Lambda}
\Lambda=\frac{l_x+l_y}{|\bar\sigma| l_x l_y}\ .
\ee
Since for large values of $\xi$, $\bar\epsilon_n\sim 1/Lambda\sim
|\xi|\lless |\xi|^2 $, the sum in\equ{exsum1} effectively never
extends to values of $n$ where the expansion of\equ{aspn} is not
justified.

We appeal to the Abel-Plana formula\cite{AP},
\bl{APf}
\sum_{n=0}^\infty \Phi(n)=\frac{\Phi(0)}{2} +\int_0^\infty dn\,
\Phi(n) +i\int_0^\infty dz \frac{\Phi(iz)-\Phi(-iz)}{\exp(2\pi
z)-1}\ ,
\ee
to evaluate the sum in\equ{exsum1} asymptotically. In the
evaluation of the first and third terms of\equ{APf}, we
approximate the exponential factor in\equ{exsum1} by unity.  We
find, using\equ{evs} that,
\bl{term1}
\frac{\Phi(0)}{2}=\frac{1}{2\sqrt{\bar\epsilon_0}}=
\left(\frac{1}{9\pi}\right)^{1/3}+O(\Lambda)\ ,
\ee
to leading order in $\Lambda$. To the same accuracy, the third
term of\equ{APf} becomes,
\bl{term3}
-2\left(\frac{2}{3\pi}\right)^\frac{1}{3}\int_0^\infty dz
\frac{{\rm Im}[(iz+\frac{3}{4})^{-\frac{1}{3}}]}{\exp(2\pi z)-1}
+O(\Lambda)\ .
\ee
To evaluate the integral over $n$ of the second term in\equ{APf},
we change the integration variable from $n$ to $\bar\epsilon_n$,
using\equ{evs}. We then find that
\bl{term2}
\int_0^\infty dn
\Phi(n)=\int_{\frac{(9\pi)^{2/3}}{4}}^\infty \frac{d\epsilon}{\pi}
\exp\left[i\frac{\epsilon^2}{4\bar\sigma} \frac{l_x+l_y}{l_x
l_y}\right]\sim \sqrt{\frac{i\bar\sigma l_x
l_y}{\pi(l_x+l_y)}}-\frac{(9\pi)^{2/3}}{4\pi}+O(\Lambda)\ .
\ee
Collecting the results and using the them in\equ{expandlo}, the
Green function at the grazing angle is found to have the
asymptotic form,
\bl{GRfinal1}
\left. G_r(0,0;\tau_c,R)\right|_{l_o=0}=
\frac{1}{2}\sqrt{\frac{k}{2\pi
i(l_x+l_y)}}+K\frac{(kR)^\frac{1}{3} e^{\pi i/3}}{4\pi i\sqrt{l_x
l_y}}+O\left(\frac{R(l_x+l_y)}{(l_x l_y)^\frac{3}{2}}\right)\ .
\ee
For large values of $kR$, the Green function at the grazing angle
is just half the direct term, with an additional
contribution\cite{const} proportional to $(kR)^\frac{1}{3}$. The
proportionality constant $K$ of the latter is
\bal{defK}
K&=&P(0)+ \frac{\pi}{(9\pi/4)^{1/3}}-
\left(\frac{9\pi}{4}\right)^\frac{2}{3} -
2\left(\frac{\pi^2}{3}\right)^\frac{1}{3}\int_0^\infty dz
\frac{2{\rm Im}(iz+\frac{3}{4})^{-\frac{1}{3}}}{\exp(2\pi z)-1}\cr
&=&P(0)-2 {\rm Re} P(0)=-P^*(0)\approx -0.966 e^{-0.00347\pi i}\ .
\ea
$(-){\rm Re} K$ differs from the coefficient $c\approx0.996$ for
the glancing contribution to the Green function of Rubinow and
Wu~\cite{const} by about 3\%; we find ${\rm Im} P(0)=-{\rm Im} K=$
to be very small, while in\cite{const} it is found to be zero. The
small discrepancy arises because the integral over a ratio of Airy
functions that defines $c$ is evaluated numerically in
\cite{const}, rather than in saddle point approximation. Our
approach is perhaps slightly more consistent -- that does not
imply that it is more accurate -- and in any event, the difference
is quite small.

It perhaps is of some interest that the correction of order
$(kR)^{1/3}$ for the glancing ray does not appear in ${\cal
G}_r(0,0;\tau_c)$. Eqs.\no{inicond} and\no{decompGr} imply that
\bal{glanceGr}
\left.{\cal G}_r(0,0;\tau_c)\right|_{\tau_o=0}&=&\frac{k}{2\pi i}
\int_0^\infty \!\!\! \frac{da}{\sqrt{l_x}}\int_0^\infty \!\!\!
\frac{db}{\sqrt{l_y}} \, \delta(a-b)\,\exp \left(\frac{i k
a^2}{2l_x}+\frac{i k b^2}{2l_y}\right)\cr
&=&\frac{1}{2}\sqrt{\frac{k}{2\pi i(l_x+l_y)}}\ ,
\ea
without further corrections.  Physically, this discrepancy can be
traced to the assumed decomposition of the radial Green function
in\equ{decompGr}. The lower bound of the integral over the radial
coordinates in\equ{decompGr} can be set to zero only for
$k=\infty$. For large but finite wave number $k$, the correction
of order $(kR)^{1/3}$ is reproduced by assuming that the
fluctuations effectively penetrate the obstacle a small distance
$d_o(k,R)$. Replacing the lower bound of the radial integrals
in\equ{glanceGr} by $-d_o(k,R)$, one finds,
\bal{glanceGrcut}
\left.{\cal G}_r(0,0;\tau_c)\right|_{\tau_o=0}&=&\frac{k}{2\pi
i\sqrt{l_x l_y}} \int_{-d_o}^\infty \!\!\! da \exp\left(\frac{i k
a^2(l_x+l_y)}{2 l_x l_y}\right)\cr &\sim
&\frac{1}{2}\sqrt{\frac{k}{2\pi i(l_x+l_y)}}-\frac{k d_o}{2\pi
i\sqrt{l_x l_y}}+ O(d_o^3)\ .
\ea
Comparison of\equ{glanceGrcut} with\equ{GRfinal1} leads to
\bl{ro}
d_o(R,k)\sim -K\frac{(k R)^{1/3}e^{\pi i/3}}{2 k}\
\ee
for $d_o(k,R)$. The effective depth of penetration thus vanishes
rather slowly,  as $k^{-2/3}$. The phase $e^{\pi i/3}$ in the
expression\equ{ro} for $d_o$ has its origin in the analytic
continuation of $R$ to $-R$, and the proportionality constant $K$
is very close to $-1$.

The true semiclassical Green function $G_r$ does not in general
separate into free- and creeping- Green functions, and
approximately reproduces the subleading asymptotic contributions
as well (to within about 3\%).

\section{Discussion, Generalization and Conclusion}
The semiclassical description of diffraction is closely associated
with extremal classical paths that satisfy Fermat's principle but
are {\it not} stationary.  Such paths arise due to the
non-holonomic constraints imposed by "obstacles". The associated
Lagrangian is given in\equ{L}. It depends on the transverse
deviations $\underline{x}_r(t),\underline{x}_\perp(t)$ from the
classical path and describes the two-dimensional motion of a
non-relativistic particle of mass $m_E=E/v^2(E)$ under the
influence of a spatially constant but in general time-dependent,
force $m_E g(t)$. The acceleration $g(t)$ is inversely
proportional to the curvature of the classical path and vanishes
for time intervals in which the classical path $\gamma_c$ is not
constrained by an obstacle. On "creeping" sections of the
classical trajectory, fluctuations are in addition restricted to
the half-space exterior to the obstacle.

Depending on the sign of the curvature, $g(t)$ is either directed
toward or away from the obstacle's surface -- corresponding to the
presence of (a generally time dependent) "floor" or "ceiling" in
an analogous 1-dimensional gravitational-like problem.  Motion in
the presence of a "floor" is relevant for "whispering-galleries",
for example, whereas the presence of a "ceiling" corresponds to
diffraction into the classical shadow of an obstacle.

Diffraction is semiclassically described by the amplitude that the
particle moves from the ceiling to the ceiling in a fixed time
interval with acceleration $g(t)$. This motion is classically
forbidden. To calculate the amplitude semiclassically, we relate
it to the one for motion from the "floor" to the "floor" with
acceleration $g(t)$. The latter problem, that of a "whispering
gallery", admits classical solutions. The amplitudes of the two
problems are related by analytic continuation in the phase $\phi$
of the curvature $R(t)\rightarrow |R(t)| e^{i\phi}$ of the
classical path.

The existence of a caustic, the "floor", that is approached by an
infinite number of classical paths poses an additional problem in
the semiclassical description. For any fixed value of $k$, there
are distinct classical paths whose actions differ by much less
than $\hbar$. The fluctuations about different classical paths
therefore cannot be separated and tend to interfere with each
other. Attempting to evaluate the path integral asymptotically by
summing the asymptotic contributions of every individual classical
path becomes highly inaccurate.

The problem was solved for the special case of a classical ray
that partially creeps along the surface of a sphere; the
asymptotic analysis of the same problem was also at the heart of
the original geometrical interpretation of
diffraction\cite{Ke62,KL70}. The curvature of this trajectory is
piecewise constant and corresponds to an acceleration $g(t)$ that
is piecewise constant in time. A path with $n$ bounces in total
time $\tau_c$ is stationary with respect to a variation of the
(conserved) "energy", $e_n$, of the particle in the region of
constant acceleration.  One thus can express the asymptotic
contribution from a classical path to given $n$ as the result of
an integral over the "energy" $e$ that is evaluated
semiclassically at the saddle point $e=e_n$. This manipulation is
consistent with the semiclassical approximation and allows one to
sum over $n$ for a given "energy" $e$ {\it before} evaluating the
remaining integral over $e$ asymptotically.

In the case of a sphere, the exact expression for the Green
function of a massless particle is known and has been evaluated
asymptotically in the umbra\cite{Ke62} of the sphere as well as in
its penumbra\cite{Pr96,const}. Quite different asymptotic forms
were found for the two cases. The procedure outlined above and
described in more detail in section~3 gives a uniform asymptotic
expansion that is valid in both regions: the exponentially
decaying terms arise from pole contributions to the final integral
over $e$, whereas power corrections (previously obtained
separately for the penumbra\cite{const}) turn out to be associated
with an end-point contribution to the integral at $e=0$.

The basic analysis through section~4 is applicable to a much more
general case, but the derivation of a relatively simple explicit
form of the asymptotic Green function,\equ{GRfinal}, was obtained
only for the spherical case. The procedure can in principle be
modified to include cases where the curvature $R(t)$ is {\it not}
piecewise constant. Because the acceleration is not constant, the
energy $e_n$ in this case is not conserved on the creeping
segment. One nevertheless may, for instance, express the action
$S_n(h_n)$ of a classical path with $n$ bounces in terms of, say,
the maximal "height" $h_n$ of the classical trajectory. It is not
difficult to verify that $h_n$ is a stationary point of $S_n(h)$
if one allows this height to vary while keeping fixed the
endpoints and number of "bounces" $n$. One can then proceed as in
section~4 and express the semiclassical approximation to the Green
function as an integral over $h$ and first sum over the number of
bounces $n$ {\it before} evaluating the integral over $h$
asymptotically. However, unlike the case considered in section~4,
the summation over $n$ generally cannot be performed in closed
form if the acceleration is not piecewise constant. One
nevertheless may sometimes be able to pick out the smallest values
of $h$ for which the sum over $n$ becomes singular and thus
(numerically) obtain the leading asymptotic approximation in the
general case.

With a view to applying the method to electromagnetic problems
with ideal conductors, we considered only Dirichlet boundary
conditions, with a change in the phase by $(-1)$ at each
reflection. Semiclassically, other idealized boundary conditions
change the phase by a different value. Furthermore, if absorption
cannot be neglected, the modulus of the reflection coefficient is
less than unity. Although the physical effect of such changes can
be dramatic, since the poles are moved, it is a great virtue of
the semiclassical approximation that neither changes in the
boundary conditions nor the inclusion of absorption alter the
procedure in any fundamental way.

We briefly comment now on two related papers. Our approach is
similar in spirit to the one of V.N.~Buslaev\cite{Bu68}. Buslaev
writes the Green function $G$ as a "continuum integral" product of
Green functions for infinitesimal time intervals and uses the
saddle point method to obtain the asymptotic form of each factor.
We do not explicitly use the product form and treat the problem by
expansion about the classical paths of extremal, but finite,
length. Our approach appears to be simpler and reduces the
diffraction problem to the propagation of a particle in a
two-dimensional space under the influence of a gravitational-like
force. However, Buslaev considers a mathematically more general
problem than we do, that is asymptotic solutions to general
parabolic equations. He also does not restrict the dimensionality
of space to three. Furthermore, our treatment of the boundary
conditions is very different from that of Budaev. McLaughlin and
Keller\cite{Mc75} note that Buslaev's result agrees with Keller's
geometrical theory of diffraction for the field of a surface
diffracted ray. An interesting alternative to these methods may be
the solution of the wave equation by analytic continuation of a
random walk proposed in ref.\cite{Bu02}.

\noindent{\bf Acknowledgements:} We thank J.B.~Keller for bringing
references\cite{Bu68} and\cite{Bu02} to our attention and for a
number of useful comments. This work was supported by the National
Science Foundation with Grant PHY-0070525. M.S. greatly enjoyed
the hospitality of New York University.

\appendix
\section{The Proportionality Factor}
To derive the relation of\equ{rel1} between the semiclassical
energy Green function $G(\bx,\by;E)$ and the Green functions
${\cal G}_{\gamma_c}$ defined in\equ{Greentau} that describes
transverse deviations from a classical path $\gamma_c$, consider
first the general case of a medium with a smooth, everywhere
differentiable index of refraction $n(\bx,E)$. In this case
classical trajectories $\gamma_c$ from $\bx$ to $\by$ are
stationary points of the action and the usual semiclassical
formalism applies. In particular, the semiclassical energy
Greenfunction is given\cite{Gu90a} by,
\bl{semigen}
G(\bx,\by;E)=\sum_{\gamma_c} \frac{1}{2\pi\hbar^2} \sqrt{D_{\gamma_c}}
\exp\left(\frac{i}{\hbar} S(\gamma_c,E)\right)\ .
\ee
For a given classical path $\gamma_c$, $D_{\gamma_c}$ is the
determinant of a $4\times 4$ matrix,
\bl{det1}
D_{\gamma_c}=\det\left(\begin{array}{c|c}
               \begin{array}{c}\\ \quad \frac{\partial^2 S(\gamma_c,E)}{\partial \bx\partial\by}\quad\\
\hfil
\end{array} &  \frac{\partial^2 S(\gamma_c,E)}{\partial\bx\partial E}\\
\hline\\
\frac{\partial^2 S(\gamma_c,E)}{\partial\by\partial E} &
\frac{\partial^2 S(\gamma_c,E)}{\partial E^2}
\end{array}\right)\ ,
\ee
with elements that are $3\times 3$, $3\times 1$, $1\times 3$ and
$1\times 1$ matrices. The form of $D_{\gamma_c}$ can be simplified
by choosing a local coordinate system in the neighborhood of the
trajectory $\gamma_c$ from $\bx$ to $\by$: the coordinate axis for
$x_\ll$ runs along the particular trajectory and the remaining
coordinates $\vec x=(x_r,x_\perp)$ are transverse to the
trajectory. Since $1/v_g$ at the endpoint of a trajectory is just
the change in the total time\equ{tott} associated with an
infinitesimal displacement of the endpoint in a direction tangent
to the trajectory, two of the second variations of the action in
\equ{det1} are given by the group velocities at the endpoints,
namely,
\begin{eqnarray}\label{dTdq}
v^{-1}_g(\bx,E)&=&-\frac{\partial{\cal T}}{\partial x_\ll} =
-\frac{\partial^2 S(\gamma_c,E)}{\partial x_\ll\partial E}\cr
v^{-1}_g(\by,E)&=&\frac{\partial{\cal T}}{\partial y_\ll} =
\frac{\partial^2 S(\gamma_c,E)}{\partial y_\ll\partial E}
\end{eqnarray}
In this coordinate system, $D_{\gamma_c}$ becomes\cite{Gu90a}
\bl{det2}
D_{\gamma_c}=\frac{1}{v_g(\bx,E)
v_g(\by,E)}\det\left(\frac{-\partial^2 S(\gamma_c,E)}{\partial
{\vec x}\partial{\vec y}}\right)\ ,
\ee
proportional to  the determinant of a $2\times 2$ matrix. Crucial
in deriving\equ{rel1} is the observation that the matrix of second
derivatives of $S$ with respect to {\it transverse} deviations
from the classical path $\gamma_c$ at the endpoints can also be
obtained by expanding $S$ to quadratic order in the transverse
deviations of the path $\gamma$ from the classical path
$\gamma_c$. We thus have that
\bl{second}
\left(\frac{-\partial^2 S(\gamma_c,E)}{\partial {\vec
x}\partial{\vec y}}\right)=\left(\frac{-\partial^2
S^{sc}_{\gamma_c}({\vec a}={\vec b}=0;\tau(\gamma_c,E))}{\partial
{\vec a}\partial{\vec b}}\right)
\ee
On the other hand, if $v(\bx,E)$ is a smooth function of the
coordinate, the classical trajectory  $\gamma_c$ is stationary and
$S^{sc}$ is quadratic in the fluctuations. In this case, the
semiclassical approximation to $\G_{\gamma_c}({\vec a},{\vec
b};\tau)$  is exact. The solution to the equation of motion for a
quadratic action $S^{sc}$ is unique and there is only one
classical trajectory $\tilde\gamma_c$ from ${\vec a}$ to ${\vec
b}$ in time $\tau$ that contributes to the two-dimensional
Greenfunction ${\cal G}_{\gamma_c}({\vec a},{\vec b};\tau)$. Van
Vleck's formula in two-dimensional space\cite{VF} then gives
\bl{greenexact}
\G_{\gamma_c}({\vec a},{\vec b};\tau)=\frac{1}{2\pi i\hbar}
\sqrt{\det\left(\frac{-\partial^2 S^{sc}_{\gamma_c}({\vec a},{\vec
b};\tau)}{\partial {\vec a}\partial{\vec
b}}\right)}\exp\left\{\frac{i}{\hbar} S^{sc}_{\gamma_c}({\vec
a},{\vec b};\tau)\right\} \ .
\ee
Note that the only classical trajectory from ${\vec a}=0$ to
${\vec b}=0$ in time $\tau(\gamma_c,E)$ is the trivial one with
vanishing transverse deviation everywhere and that therefore
\bl{equiv}
S(\gamma_c,E)=S^{sc}_{\gamma_c}({\vec a}={\vec b}=0;
\tau(\gamma_c,E))\ .
\ee
Comparing\equ{greenexact} with\equ{semigen} for
$\tau=\tau(\gamma_c,E)$ and ${\vec a}={\vec b}=0$ and using
Eqs.\no{equiv},\no{second} and\no{det2},  one arrives at the
relation\equ{rel1}.

Strictly speaking the validity of\equ{rel1} has only been shown
for a smooth, everywhere differentiable phase velocity $v(\bx,E)$.
In this case the classical trajectories $\gamma_c$ are stationary
points of the action and the semiclassical expressions we employed
are valid. We argue that\equ{rel1} also holds when $v(\bx,E)$ is a
step function that vanishes within ${\cal V}$, because \equ{rel1}
does not explicitly depend on the spatial dependence of $v(\bx,E)$
and involves only the group velocities at the endpoints of the
classical trajectories. \equ{rel1} does not depend on the nature
of the obstacle and the discontinuous spatial dependence of the
phase velocity of interest can be considered as a limiting case.
One similarily can  argue that because\equ{rel1} holds for
endpoints $\bx$ and $\by$ connected by classical trajectories that
do not touch the obstacle, the relation should by continuity
remain valid when one of the endpoints lies in the classical
shadow region of the other.

\section{Proof that ${\cal D}_n(\xi)\rightarrow D_n(\xi)$ for
$n\rightarrow\infty$} It is instructive to see that the
diffraction amplitudes defined in Eqs.\no{defD} and\no{Dsemi}
coincide in the limit of large $n$ for any  fixed real value of
$\xi$. For large $n\sim\infty$ the zero's of the Airy function
$\epsilon_n$ approach $\bar\epsilon_n$ defined by\equ{evs}; in
particular, $\epsilon_n$ becomes arbitrarily large for
$n\rightarrow\infty$. {\vskip0.1truecm\begin{center}
\epsfig{figure=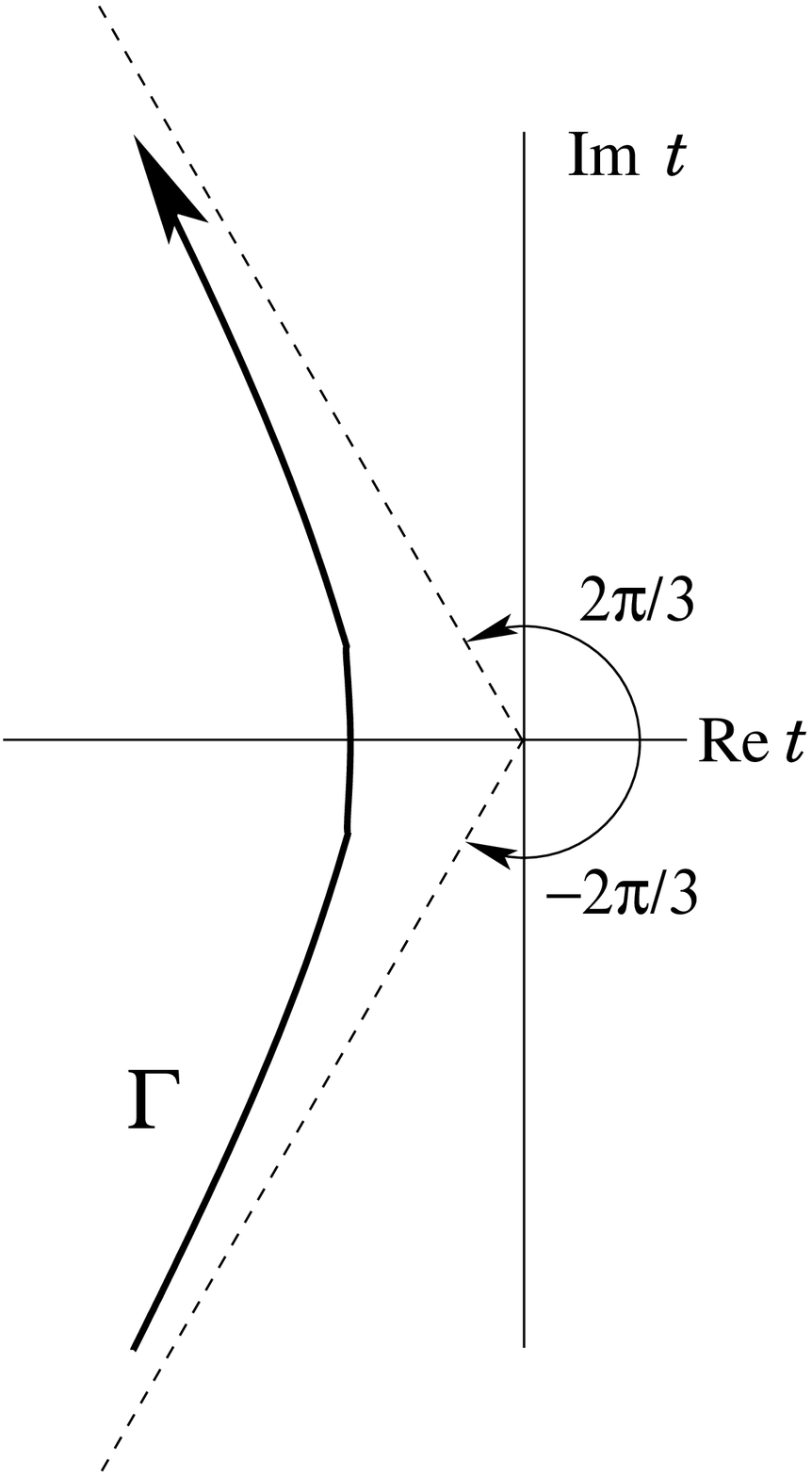,height=5.0truecm}\end{center}\nobreak

{\small\noindent Fig.~5: The contour $\Gamma$ in the
representation for the Airy-function $\Ai(z)$ of\equ{AiryAi}. For
large values of $|t|$,  $\Gamma$ approaches the asymptotes with
arg~$t=\pm 2\pi/3$ shown as dashed lines.}

The Airy function can be represented by the contour
integral\cite{AS}
\bl{AiryAi}
\Ai(z)=\frac{1}{2\pi i}\int_\Gamma dt e^{t z -t^3/3}\ ,
\ee
where the contour $\Gamma$ is sketched in Fig.~5. The expression
for ${\cal D}_n(\xi)$ of\equ{defD} then becomes
\bl{DD}
{\cal D}_n(\xi)=\frac{1}{2\pi i
|\Ai^\prime(-\epsilon_n)|}\int_0^\infty d\rho\int_\Gamma dt
e^{i\frac{\rho^2}{4}-\frac{t^3}{3}+t\rho\sqrt{\xi}-t\epsilon_n}\ .
\ee
For large values of $\epsilon_n$ the saddle-point approximation to
the integral of\equ{DD} is accurate.  The two saddle points of the
integrand are located at
\bl{saddlepm}
\bar\rho_n^\pm = 2\sqrt{\xi} \bar p_n^\pm(\xi)\ , \ \
\bar t_n^\pm =-i \bar p_n^\pm(\xi))\ ,
\ee
with
\bl{defppm}
\bar p_n^\pm(\xi)=\pm\sqrt{\epsilon_n+\xi^2}-\xi\ .
\ee
For large values of $n$, $\bar \rho_n^-$ is negative and does not
contribute to the saddle point approximation of the integral. The
path of integration can, however, be deformed to pass over the
other stationary point at $(\bar \rho_n^+,\bar t_n^+)$. The
integral of\equ{DD} in saddle point approximation thus becomes,
\bl{saddleDD}
{\cal D}_n(\xi)\sim\frac{1}{
|\Ai^\prime(-\epsilon_n)|(\epsilon_n+\xi^2)^{1/4}}\exp i\left[
\frac{2}{3} \left(\bar p_n^+(\xi)\right)^3+\xi \left(\bar
p_n^+(\xi)\right)^2\right]\ .
\ee
One finally arrives at the expression of\equ{Dsemi} for $D_n(\xi)$
by using the asymptotic formula for the Airy function.
\equ{boundAi} implies that for large $n$,
\bl{asAiprime}
\epsilon_n\sim\bar\epsilon_n,\ \bar p_n^+(\xi)\sim \bar p_n(\xi)\
\ {\rm and}\ |\Ai^\prime(-\epsilon_n)|\sim
\bar\epsilon_n^{1/4}/\sqrt{\pi}\ ,
\ee
where $\bar\epsilon_n$ and $\bar p_n(\xi)$ are defined
in~Eqs.\no{evs} and\no{barq}, respectively. Although the
amplitudes ${\cal D}_n(\xi)$ and $D_n(\xi)$ coincide for large
values of $n$, this does not mean that the contribution from the
poles of\equ{semsing}, by itself,  gives the correct asymptotic
expansion of the Green function. The asymptotic expansion of the
Green function is in fact dominated by low-$n$ terms, and we
indeed obtained asymptotic corrections to the pole contributions
in the semiclassical approximation.

\section{Asymptotics of $P(z)$}
The function $P(z)$ defined by\equ{defP} may be written as a
McLaurin series,
\bl{PLaurin}
P(z)=\sum_{n=1}^\infty\frac{(-iz)^{n-1}}{\Gamma(n)} P_n\ ,
\ee
with coefficients $P_n$ given by
\bl{Pns}
P_n=\sqrt[3]{18}\int_{{\cal C}_{sd}}  \xi^{2 n-1} d\xi  \,
\frac{ie^{-i\xi^3}}{1+i
e^{-i\xi^3}}=-\sqrt[3]{18}\sum_{k=1}^\infty (-i)^k\int_{{\cal
C}_{sd}}  \xi^{2 n-1} d\xi  \, e^{-ik\xi^3} \ .
\ee
For given $k$, the integral for the contour ${\cal C}_{sd}$ of
Fig.~4 can be performed by making the substitution
\bl{rh}
\xi=\sqrt[3]{\rho/k} e^{-i\pi/6}\ ,
\ee
where $\rho$ is real, and using the definition\no{Li}. We then
have
\bal{Pns1}
P_n&=&-\frac{\sqrt[3]{18}}{3} e^{-i\pi n/3} \Gamma(2
n/3)\sum_{k=1}^{\infty}\frac{(-i)^k}{k^{2n/3}}\cr &=& i
\left(2i/3\right)^{1/3} e^{-\frac{i\pi (n-1)}{3}}\Gamma(2n/3) {\rm
Li}_{2 n/3}(-i)\ .
\ea
Eqs.\no{PLaurin} and\no{Pns1} give the convergent series
of\equ{seriesP}. Note that the series in\equ{seriesP} uniquely
defines $P(z)$ in the whole complex plane but for $|z|\gg 1$, the
accurate evaluation of $P(z)$ requires a sizable number of terms
and is numerically not very efficient.

For $|z|\gg 1$, terms of order $\xi^3$ in the exponent may be
ignored to leading order of the steepest descent method and the
integral of\equ{defP} approaches the asymptotic form in\equ{asP},
\bl{stdP} P(z\gg 1)\sim i\sqrt[3]{18}\int_{{\cal
C}_{\rm sd}} \xi d\xi \frac{e^{-i z
\xi^2}}{1+i}=\frac{3^{2/3}}{2^{7/6} z} e^{-i\pi/4}.
\ee
For real $x$, $P(x)$ is a rather slowly changing function of $x$
that smoothly interpolates between $P(0)$ and $P(x\gg 1)$. For
real positive $x$, $P(x)$ is numerically approximated by the
rational function
\bl{Papprox}
P(x)\approx\frac{0.7(1-i)}{0.7(1-i)+x}\
\ee
to within about $10\%$.


\begin{thebibliography}{99}
\bibitem{fnote1} Motion on a helix with constant speed
$v=\sqrt{v_z^2+(R_{\rm cyl}\omega)^2}$ can be parameterized by
$\bx_{\rm helix}(t)= (R_{\rm cyl} \cos (\omega t), R_{\rm
cyl}\sin(\omega t), v_z t)$, where $\omega$ is the constant
clockwise angular velocity around the cylinder of radius $R_{\rm
cyl}$ and $v_z$ is the velocity component parallel to the axis of
the helix. ($R_{\rm cyl}$ is the curvature of the helical path
only if $v_z=0$.) Using\equ{locbasis} to obtain the corresponding
local basis vectors and evaluating\equ{skew}, one finds that
$\alpha=-\omega v_z/v$ in this case.
\bibitem{fnote2} The whole semiclassical
approximation is only of pedagogical value in the case of a
sphere, since the exact spectral representation of the Green
function of a massless particle excluded from a spherical region
is known.
\bibitem{fnote3} A "bounce" here is analogous to an
 "instanton" solution in Quantum Field Theory in the sense that
it too is a time-dependent and topologically stable solution of
the classical equation of motion.
\end{thebibliography}
\end{document}